\pdfoutput=1

\documentclass[letterpaper,twocolumn,10pt]{article}
\usepackage{usenix2019_v3,epsfig}
\usepackage[bf,small]{caption} 

\usepackage[normalem]{ulem}

\usepackage{url}
\usepackage{graphicx}
\usepackage{listings}
\usepackage{epstopdf}
\usepackage{color}
\usepackage{subfigure}
\usepackage{xspace}
\usepackage{latexsym}
\makeatletter
\newif\if@restonecol
\makeatother

\usepackage[ruled,vlined,linesnumbered]{algorithm2e}
\usepackage{algorithmic}
\usepackage{enumerate}

\usepackage{makecell}

\usepackage{fontawesome}

\usepackage{wrapfig}
\usepackage{setspace}
\usepackage{blindtext}
\usepackage{amsfonts}
\usepackage{multirow}
\usepackage{pifont}
\usepackage{mathrsfs}
\usepackage{mathtools}
\usepackage{relsize}
\usepackage{ulem}
\usepackage{comment}
\usepackage{footnote}
\usepackage{tikz}
\usetikzlibrary{calc}
\makesavenoteenv{tabular}
\makesavenoteenv{table}

\newif{\ifSubmit}
\newif{\ifFinal}
\newif{\ifDraft}
\Drafttrue

\ifSubmit
\newcommand{\alicomment}[1]{}
\newcommand{\yuecomment}[1]{}
\newcommand{\vcomment}[1]{}
\else
\newcommand{\alicomment}[1]{\noindent\textcolor{magenta}{\bf Ali: #1}}
\newcommand{\yuecomment}[1]{\noindent\textcolor{red}{\bf Yue: #1}}

\newcommand{\vcomment}[1]{\textcolor{green}{\textbf{Vasily: #1}}}
\fi

\newcommand{\proj}{\textsc{InfiniCache}}

\usepackage{booktabs}
\usepackage{siunitx}

\usepackage{zref}
\usepackage{cleveref}
\crefformat{section}{\S#2#1#3}

\microtypecontext{spacing=nonfrench}

\usepackage{balance}

\usepackage[numbers]{natbib}
\newcommand*\circled[1]{\tikz[baseline=(char.base)]{
            \node[shape=circle,draw,inner sep=.3pt] (char) {#1};}}

\begin{document}

\title{\Large {\proj}: Exploiting Ephemeral Serverless Functions\\ to Build a Cost-Effective Memory Cache\thanks{This is the preprint version of a paper published in USENIX FAST 2020.}}

\author{
	{\rm Ao Wang$^1$\thanks{These authors contributed equally to this work.}, Jingyuan Zhang$^{1\dagger}$, Xiaolong Ma$^{2}$, Ali Anwar$^{3}$, Lukas Rupprecht$^{3}$, Dimitrios Skourtis$^{3}$, }\\
	{\rm  Vasily Tarasov$^{3}$, Feng Yan$^{2}$, Yue Cheng$^1$}\\
	{{{$^{1}$}George Mason University} ~~~{$^{2}$}{University of Neveda, Reno} ~~~{$^{3}$}{IBM Research--Almaden}}\\
}

\maketitle

\pagestyle{plain}

\begin{abstract}
Internet-scale web applications are becoming increasingly storage-intensive and rely heavily on in-memory object caching to attain required I/O performance. We argue that the emerging serverless computing paradigm provides a well-suited, cost-effective platform for object caching. 
We present {\proj}, a \emph{first-of-its-kind} in-memory object caching system that is completely built and deployed atop ephemeral serverless functions. {\proj} exploits and orchestrates serverless functions' memory resources to enable elastic pay-per-use caching. {\proj}'s design combines erasure coding, intelligent billed duration control, and an efficient data backup mechanism to maximize data availability and cost effectiveness while balancing the risk of losing cached state and performance. We implement {\proj} on AWS Lambda and show that it: (1)~achieves $31$ -- $96\times$ tenant-side cost savings compared to AWS ElastiCache for a large-object-only production workload, (2)~can effectively provide $95.4\%$ data availability for each one hour window, and (3)~enables comparative performance seen in a typical in-memory cache.

\end{abstract}
\vspace{-8pt}
\section{Introduction}
\label{sec:intro}
\vspace{-6pt}

Internet-scale web applications are becoming increasingly important as they offer many useful services to the end users. Examples range from social networks~\cite{haystack_osdi10} that serve billions of photo and video files every day to hosted  container image repositories such as Docker Hub~\cite{dockerhub}. These web applications typically require a large storage capacity for the massive amount of data they must store.
For instance, Docker Hub hosts over $2.6$ million container images, and Facebook generates 4 PB of data daily~\cite{fb_data}.

Cloud object stores (e.g., Amazon S3, Google Cloud Storage, OpenStack Swift, etc.) have become the first choice
for serving the simple object {\small\texttt{GET}}/{\small\texttt{PUT}} requests of these storage-intensive web applications.
To improve request latencies for better user experience, cloud object stores are typically being used in combination with networked, lookaside In-Memory Object Caches (IMOCs) such as Redis~\cite{redis} and Memcached~\cite{memcached}. 
Serving requests from an IMOC is much faster than serving them directly from a backing object store. 
However, due to the high cost of main memory, IMOCs are largely used only as a small cache for buffering small-sized objects that range in size from a few bytes to a few KBs~\cite{fbkvs_sigmetrics12}. 
Caching large objects (i.e., objects with sizes of MBs--GBs) is believed to be relatively inefficient in an IMOC
as large objects consume significant memory capacity and network bandwidth. This either causes cache churn with evictions of many small objects that would be reused soon if the cache is too small, or incurs high cost for larger cache sizes.

Large object caching has been demonstrated to be effective and beneficial in cluster computing~\cite{eccache_osdi16, spcache_sc18, alluxio_socc14, pacman_nsdi12}. 
To verify that these benefits also apply to web applications, we analyzed production traces from an IBM Docker registry~\cite{docker_fast18} and identified two key properties for large objects:
(1) large objects are heavily reused with strong data locality and are accessed less frequently than small ones, and (2) achieving a fast access speed for large objects is critical for system performance though it does not require as stringent a service level objective (SLO) as that for small objects, the latter of which demands sub-millisecond latencies. 
These properties suggest that web applications can benefit from large object caching, which state-of-the-art IMOCs currently do not provide. 

\begin{figure*}[t]
\begin{center}
\subfigure[Object size.] {
\includegraphics[width=.235\textwidth]{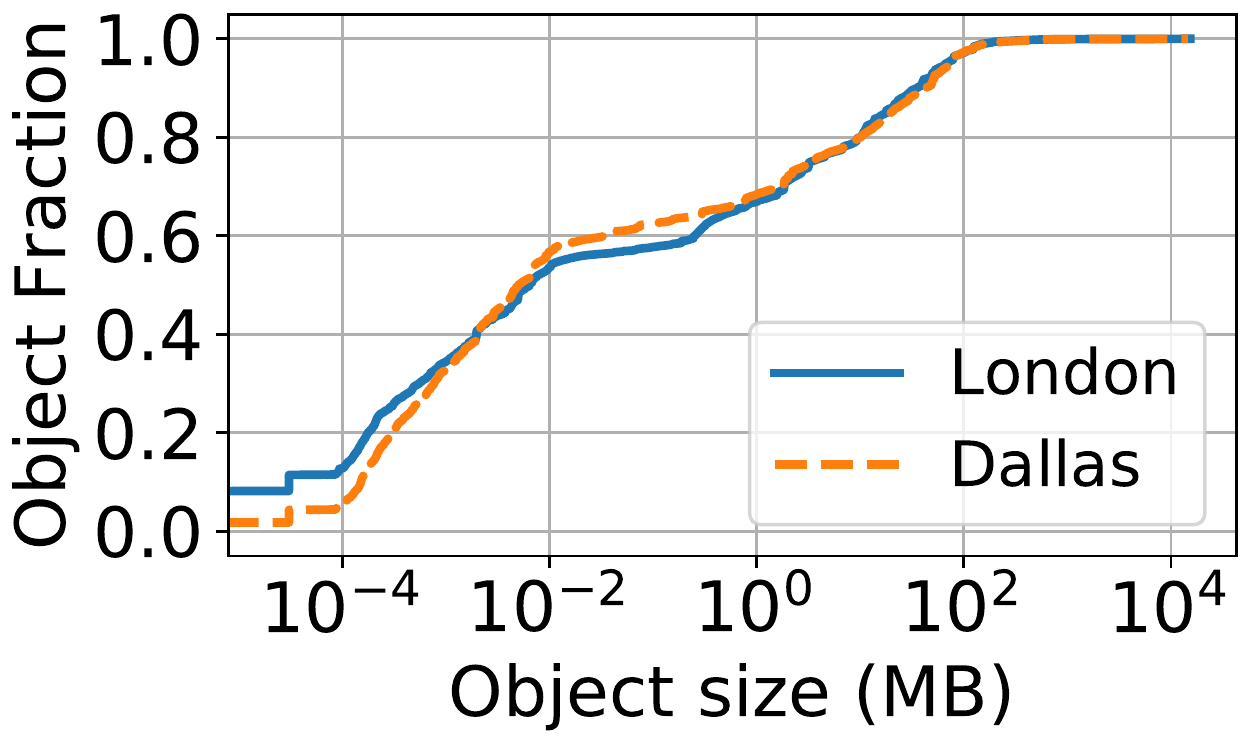}
\label{fig:obj_sz}
}
\hspace{-10pt}
\subfigure[Object footprint.] {
\includegraphics[width=.235\textwidth]{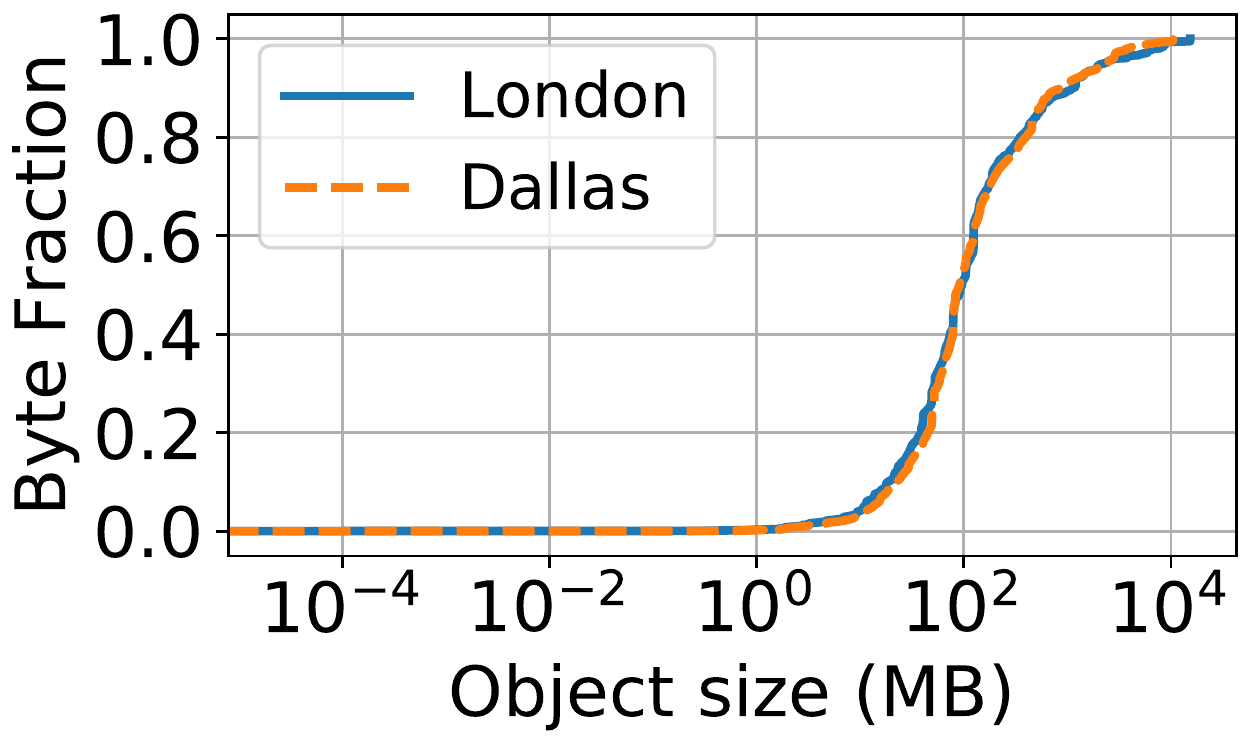}
\label{fig:obj_frac}
}
\hspace{-10pt}
\subfigure[Access count for obj. $>$ 10 MB.] {
\includegraphics[width=.235\textwidth]{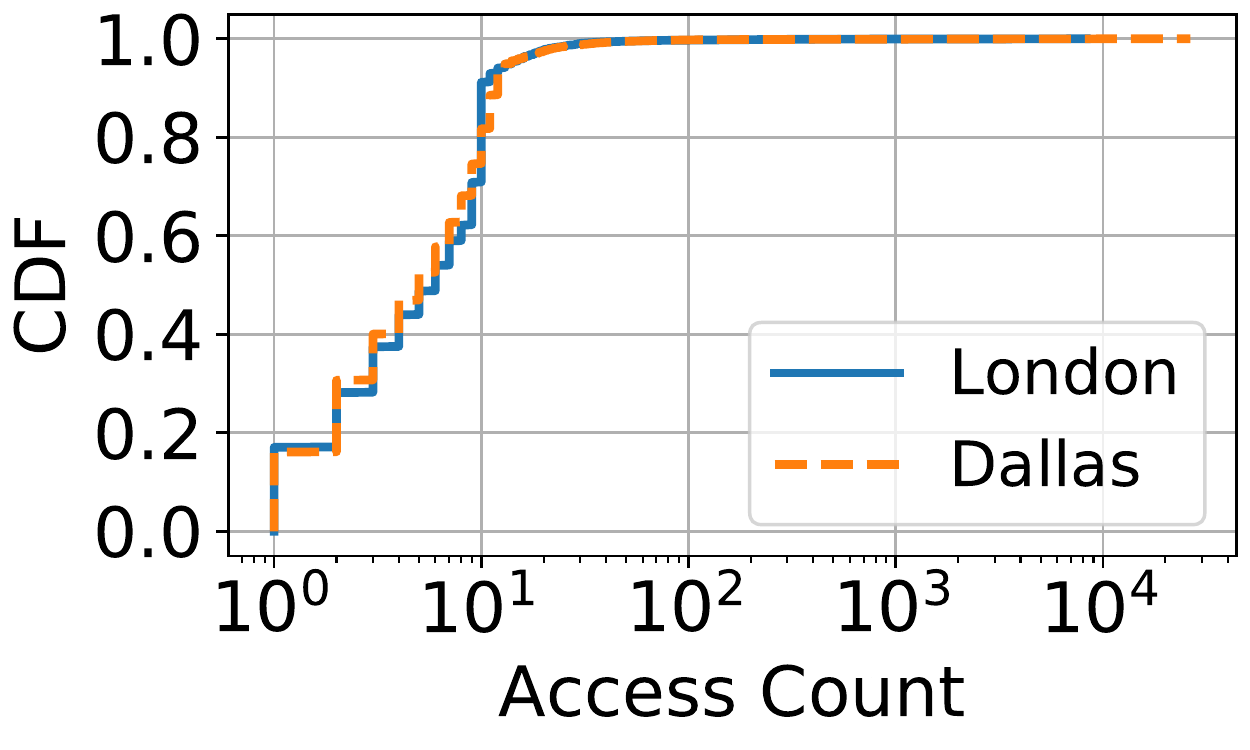}
\label{fig:obj_freq}
}
\hspace{-10pt}
\subfigure[Reuse interval for obj. $>$ 10 MB.] {
\includegraphics[width=.235\textwidth]{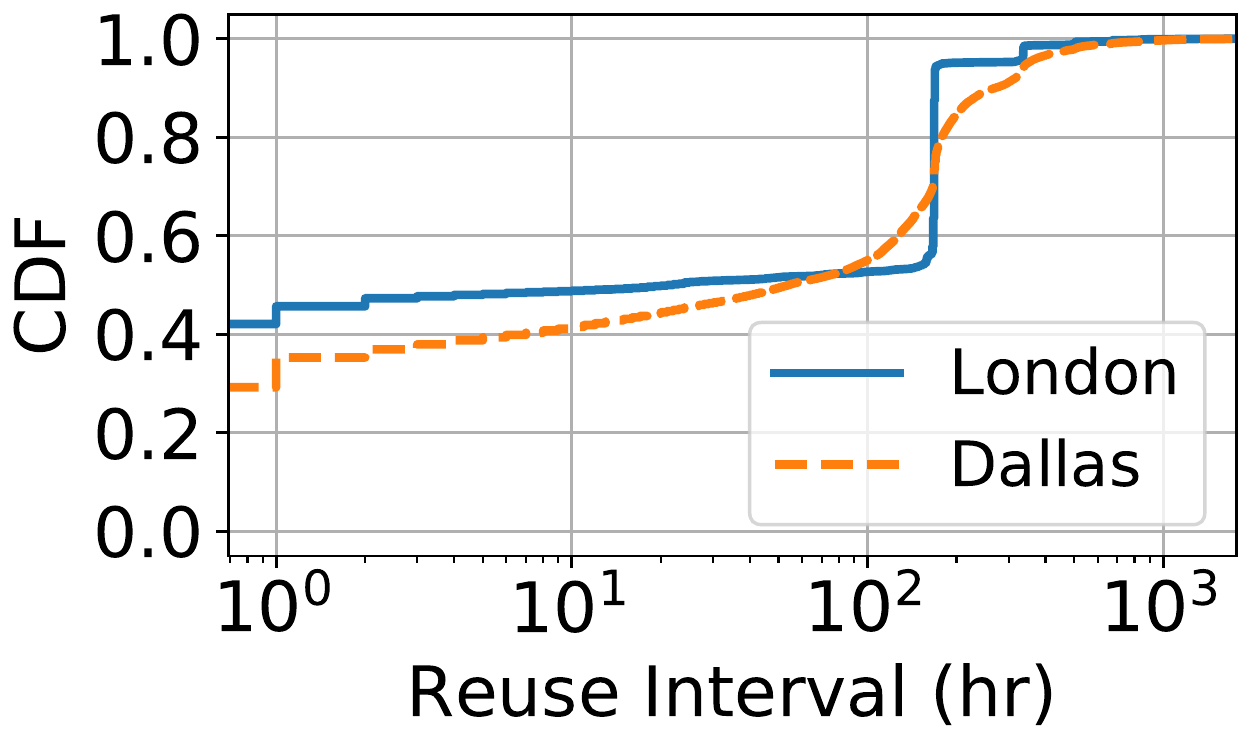}
\label{fig:obj_dist}
}
\vspace{-10pt}
\caption{Characteristics of object sizes and access patterns in the IBM Docker registry production traces. }
\label{fig:wl_char}
\end{center}
\vspace{-25pt}
\end{figure*}

The emerging serverless computing paradigm (cloud function services, or Function-as-a-Service (FaaS))~\cite{berkeley_serverless} introduces a new way of building and deploying applications, in which the service providers take care of resource scaling and management. Developers can thus focus on developing the function logic without managing servers.
Popular uses of serverless computing today are event-driven and stateless applications such as web/API serving and batch ETL (extract, transform, and load)~\cite{serverless_usecases}.
However, we find that serverless computing can also provide a potential cost-effective solution for resolving the tension between small and large objects in memory caching. 

We demonstrate how to build an IMOC as a serverless application. A serverless application is structured as a collection of cloud functions. A function has memory that can be used to store objects that are needed during its execution. We use this memory to store cached objects. Functions are executed on demand. In our serverless IMOC, the functions are invoked by the tenant to access the cached objects. FaaS providers cache invoked functions and their state so in-memory objects are retained between function invocations. This provides a sufficient lifetime for cached objects. Providers only charge tenants when a function is invoked, in our case, when a cached object is accessed. Thus the memory capacity used to cache an object is billed only when there is a request hitting that object. \emph{Our serverless IMOC reduces the tenants' monetary cost of memory capacity compared to other IMOCs that charge for memory capacity on an hourly basis whether the cached objects are accessed or not}.

Utilizing the memory of cloud functions for object caching introduces non-trivial challenges due to the limitations and constraints of serverless computing platforms: Cloud functions have limited resource capacity (e.g., 1 CPU, up to several GB memory, and limited network bandwidth) with strict network communication constraints (e.g., no inbound TCP connection); providers may reclaim a function and its memory at any time, creating a risk of loss of the cached data.

We present {\proj},
a cost-effective in-memory object cache that exploits and orchestrates serverless cloud functions. 
{\proj} synthesizes a series of  techniques into a holistic design to overcome the aforementioned challenges and to achieve high performance, cost effectiveness, scalability, and fault tolerance.
{\proj} leverages erasure coding to: (1) provide fault tolerance against data loss due to function reclamation by the service provider;
(2) improve performance by utilizing the aggregated network bandwidth of multiple cloud functions in parallel; and (3) use redundancy to handle tail latencies caused by straggling functions. 
{\proj} implements function orchestration policies that improve reliability while lowering cost.
Specifically, {\proj} implements a lightweight data backup mechanism in which a cloud function periodically performs delta synchronization (delta-sync) with a \emph{clone} of itself so as to minimize the chances that a reclaimed function causes a data loss.

\vspace{-8pt}
\begin{itemize}
    \item Identify the opportunities and challenges of serverless function-based object caching by performing a long-term analysis of the internal mechanisms of a popular serverless computing platform (AWS Lambda~\cite{lambda}).
    \vspace{-8pt}
    \item Design and implement {\proj}, the very first in-memory object caching system powered by ephemeral and ``stateless'' cloud functions.
    \vspace{-8pt}
    \item Provide an analytical model of {\proj}'s fault tolerance mechanism built using erasure coding and periodic delta-sync techniques.
    \vspace{-8pt}
    \item Perform an extensive evaluation using both microbenchmark and production workloads. Experimental results show that {\proj} achieves performance comparable to ElastiCache for large objects and improves the cost effectiveness of cloud IMOCs by $31$ -- $96\times$.
\end{itemize}

\vspace{-15pt}
\section{Background and Motivation}
\label{sec:moti}
\vspace{-8pt}

Large-scale web applications have increasingly complex storage workload characteristics. Many modern web applications utilize a microservice architecture, which consists of hundreds to thousands of microservice modules~\cite{deathstar_asplos19}. Different modules exhibit different object size distributions and request patterns. For example, a Docker image registry service uses Redis to store small-sized container metadata (i.e., manifests), and an object store 
to store large-sized container images~\cite{docker_fast18, bolt_cloud19}.
While in-memory caching has been extensively studied in the context of large-scale web applications focusing on small objects, cloud cache management for large objects remains poorly explored and poses further challenges. 

\vspace{-10pt}
\subsection{Large Object Caching}
\label{sec:large-object-caching}
\vspace{-5pt}

To obtain a better understanding of large object caching, we analyze production traces from an IBM Docker registry collected in 2017 from two datacenters (one in London, UK, and the other in Dallas, US)~\cite{docker_fast18}. The goal is to reveal patterns that enable us to make realistic assumptions for the design of {\proj}.

\vspace{-12pt}
\paragraph{Extreme Variability in Object Size.}
We first analyze the object size distributions. 
As shown in Figure~\ref{fig:obj_sz}, we find that object sizes span over nine orders of magnitude, and that more than $20\%$ of objects are larger than 10 MB in size. This observation highlights the extreme variability and heterogeneity of real-world object store workloads, which further increases the complexity of cloud IMOC management.

\vspace{-12pt}
\paragraph{Tension between Small and Large Objects.}
Efficiently managing both small and large objects in an IMOC is challenging due to two performance-cost tradeoffs. First, with limited cache capacity, large objects occupy a large amount of memory and would cause evictions of many small objects that might be reused in the near future, thus hurting performance.
This is evidenced by Figure~\ref{fig:obj_frac}, where large objects (with size larger than 10 MB) occupy more than $95\%$ of the total storage footprint.
Second, large object requests typically consume significant network bandwidth resources, which may inevitably affect the latencies of small objects. 

On one end, to prevent large objects from consuming too much memory and starving small object requests, an object size threshold is defined to not admit objects larger than the threshold~\cite{adaptsize_nsdi17, varnish}.
On the other end, system administrators can simply provision more memory (and thus more servers) to increase the capacity of the cache. However, this would increase the total cost of ownership (TCO) with reduced resource utilization. 
In fact, according to our analysis of the production Docker registry workloads, for the busiest deployment among seven datacenters, the average throughput of requests with object sizes greater than 10MB is below $3,500$ {\small\texttt{GETs}} per hour.

\vspace{-12pt}
\paragraph{Caching Large Objects Matters.}
While large object caching is challenging, it can provide significant benefit as large object workloads exhibit strong data locality. Figure~\ref{fig:obj_freq} plots the access frequency distribution for all objects larger than 10 MB. About $30\%$ of large objects are accessed at least 10 times, and the object popularity shows a long-tail distribution, with the most popular objects absorbing more than $10^4$ accesses. Figure~\ref{fig:obj_dist} shows the temporal reuse patterns of the large object workloads. Around $37\%$--$46\%$ large objects are reused within 1 hour since the last time they were accessed. The strong temporal locality patterns underscore the benefit for caching large objects for web applications.

\vspace{-10pt}
\subsection{Building a Memory Cache on Cloud Functions: Opportunities and Challenges}
\label{subsec:challenges}
\vspace{-3pt}
The above observations lead to an important question to the storage system designers and cluster administrators: \emph{can we build a new cloud caching model that relieves the tension between performance and cost while serving large objects in a cost-effective manner?}
We argue that what is missing is a truly elastic cloud storage service model that charges tenants in a request driven mode instead of capacity usage, which the emerging serverless computing naturally enables, with the following desirable properties:

\vspace{-12pt}
\paragraph{Pay-Per-Use Pricing:}
FaaS providers (including AWS Lambda~\cite{lambda}, Google Cloud Functions~\cite{google_func}, Microsoft Azure Functions~\cite{azure_func}, and IBM Cloud Functions~\cite{ibm_func}) charge users at a fine granularity -- for example, AWS Lambda bills on a per-invocation basis (\$$0.02$ per 1 million invocations) and charges (CPU and memory bundle) resource usage by rounding up the function's execution time to the nearest $100$ milliseconds with a rate of $\$0.0000166667$ per second for each GB of RAM. Note the function startup cost is not billed, and does not count for its execution time.
Large object IMOC workloads can take advantage of this fine-grained pay-as-you-go pricing model to keep the tenant's monetary costs low.

\vspace{-12pt}
\paragraph{Short-Term Caching:}
More importantly, FaaS providers keep  functions ``warm'' by caching their state in memory for a short period of time to mitigate the ``cold-start'' penalty\footnote{``Cold start'' refers to the first-ever invocation of a function instance.}~\cite{peeking_atc18, sock_atc18, sand_atc18}. Functions that are not invoked for a while can be reclaimed by the provider, and the state stored in the functions is lost. The duration of the ``warm'' period may vary (ranging from tens of minutes to longer than 6 hours as observed in \cref{subsec:aws-lambda-properties}) for AWS Lambda, and largely depends on how frequently the Lambda function gets invoked.

Ideally, a cloud tenant can leverage the above properties naturally enabled by a FaaS provider to build an opportunistic IMOC on a serverless platform. As such, a naive design would simply invoke a cloud function and store objects into the function's memory until the function is reclaimed by the provider, and then re-insert the objects into a new function.

This approach is appealing for several reasons. First and foremost, it inherently redefines the pay-as-you-go pricing model in the context of storage (in our case memory cache storage) by realizing a new form of memory elasticity --- the memory capacity used to cache an object is billed only when there is a request hitting that object. This significantly differentiates the proposed cache model against conventional cloud storage or cache services, which start charging tenants for capacity usage whenever the capacity has been committed in use. 
Second, it offers a virtually infinite (yet cheap) short-term capacity, which is advantageous for large object caching, since the tenants can invoke many cloud functions
but have the provider pay the cost of function caching\footnote{FaaS providers essentially pay for the cost of storing the objects, while the tenants pay for the function invocations and function duration.}.

However, FaaS providers place limits on the use of cloud resources to simplify resource management, which introduces challenges in building a stateful cache service atop stateless cloud functions.
Take AWS Lambda for example --- 
each Lambda function comes with a limited CPU and memory capacity;
tenants can choose a memory amount between 128MB and 3008MB in 64MB increments. Lambda allocates CPU power linearly in proportion to the amount of memory configured, capped by $1.7$ cores. 
Each Lambda function can run at most 900 seconds (15 minutes) and will be forcibly returned when the function times out. 
In addition, Lambda only allows outbound TCP network connections and bans inbound connections and UDP traffic, meaning a Lambda function cannot be used to implement a server, which is necessary for stateful applications such as IMOC. However, once an outbound TCP connection is established, it can be used to issue (multiple) requests to the function.
Another limitation that plagues the performance of serverless applications is the lack of quality-of-service (QoS) control. As a result, functions suffer from straggler issues~\cite{locus_nsdi19}.
\emph{Therefore, an ideal IMOC built atop cloud functions must provide effective workaround solutions to all the above challenges.}
\vspace{-12pt}
\section{{\proj} Design}
\label{sec:design}
\vspace{-5pt}

\begin{figure}[t]
\begin{center}
\includegraphics[width=0.4\textwidth]{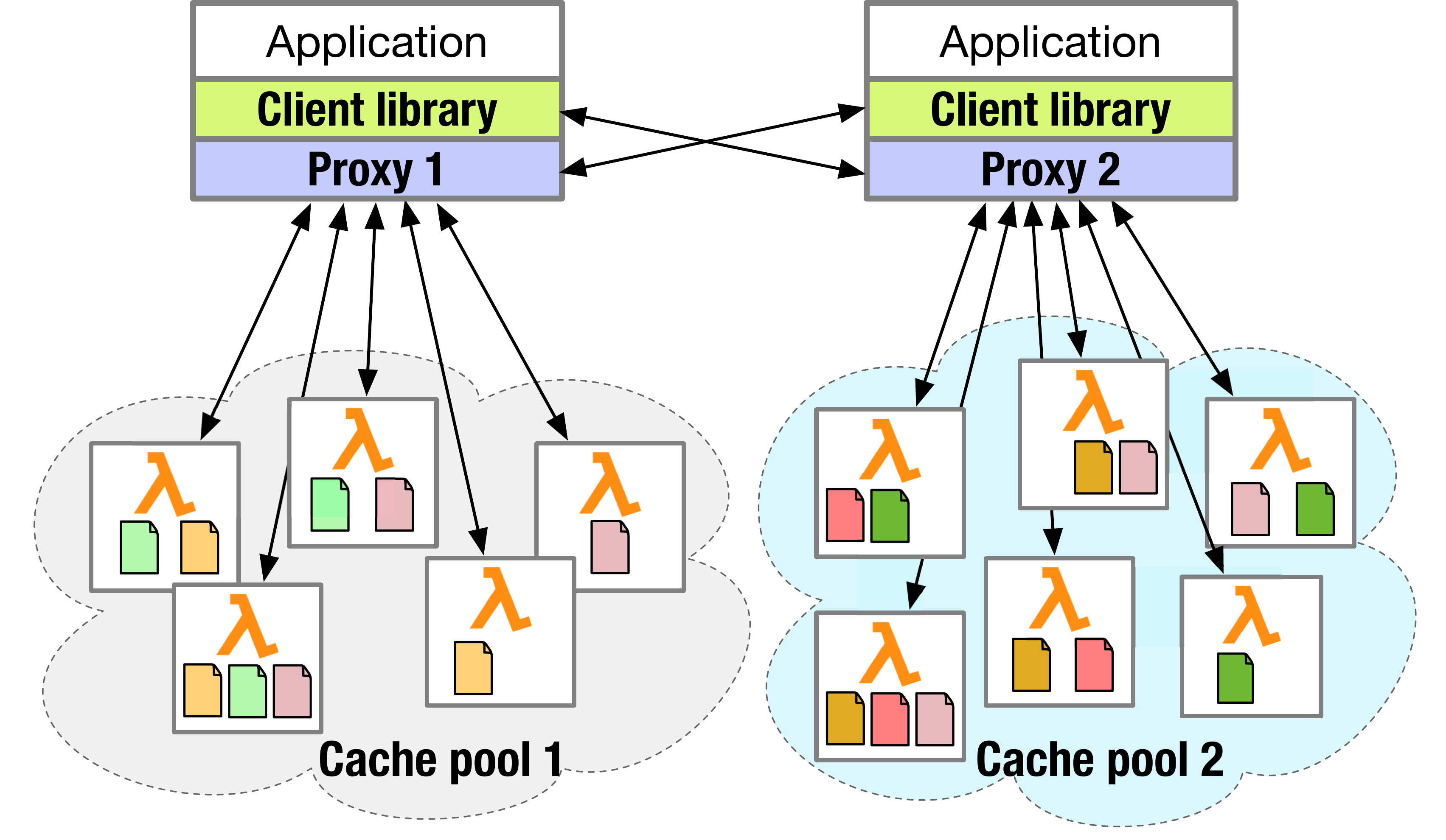}
\vspace{-10pt}
\caption{{\proj} architecture overview. 
Icon \faFile\ denotes EC-encoded object chunks. Chunks with same color belong to the same object.
}
\label{fig:arch}
\vspace{-25pt}
\end{center}
\end{figure}

{\proj} has three components: an {\proj} client library, a proxy,
and a Lambda function runtime used to implement cache nodes\footnote{We use Lambda cache node and Lambda function (runtime) interchangeably in different contexts.}. As shown in Figure~\ref{fig:arch}, an {\proj} deployment consists of a cluster of Lambda cache nodes, which are logically partitioned and managed by multiple proxies. Each proxy orchestrates a \emph{Lambda cache pool}. Applications interact with {\proj} via a client library that is responsible for cache invalidation upon an overwrite and cache insertion upon a read miss assuming a read-only, write-through cache; the client library encodes and decodes the objects using erasure coding (EC) and interfaces with a proxy serving as a rendezvous that streams the EC-encoded object chunks between a client library and the Lambda nodes.

{\proj} introduces a proxy primarily because a Lambda node cannot run in server mode due to banned inbound connections. Thus a client library has to rely on an intermediate server (the proxy) for accepting connection requests from Lambda nodes.
In {\proj}, the client library and proxy are logically separated as they have clearly partitioned functionality, but in deployment they can be physically co-located on the same machine. To enable data sharing across different Lambda cache pools, a client can communicate with any proxy (see Figure~\ref{fig:arch}).

\vspace{-8pt}
\subsection{Client Library}
\label{subsec:client}
\vspace{-5pt}

\begin{figure}[h]
\begin{center}
\vspace{-5pt}
\includegraphics[width=0.4\textwidth]{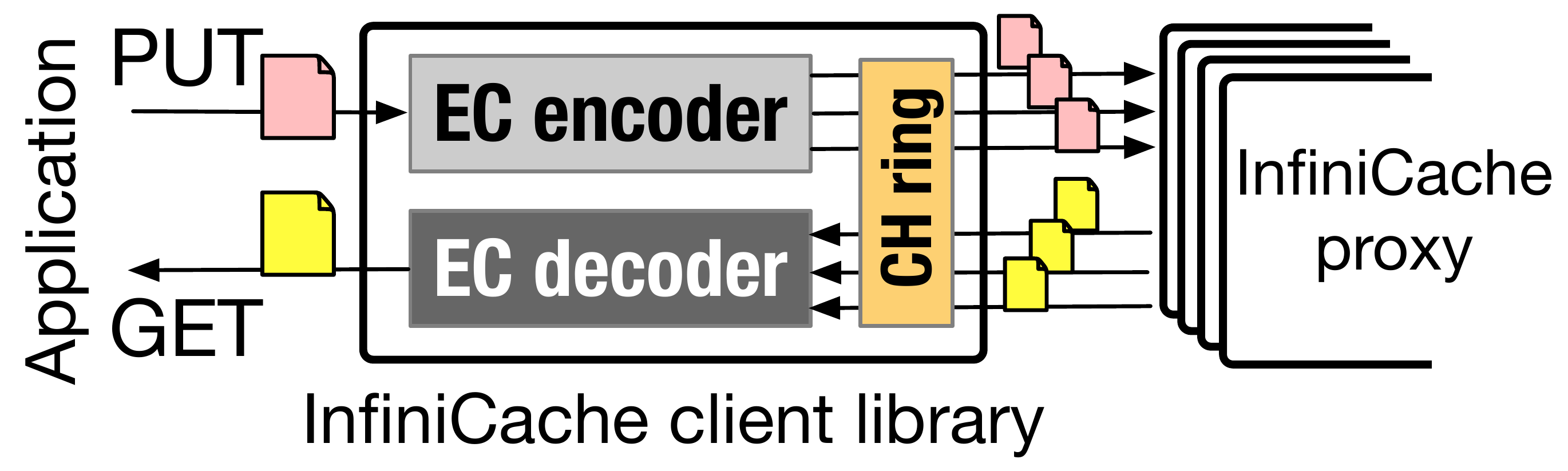}
\vspace{-7pt}
\caption{{\proj} client library ({\small\texttt{CH}}: consistent hashing).}
\label{fig:client}
\vspace{-20pt}
\end{center}
\end{figure}

{\proj}'s client library exposes to the application a clean set of {\small\texttt{GET(key)}} and {\small\texttt{PUT(key, value)}} APIs (see Figure~\ref{fig:client}). The client library is responsible for: (1) transparently handling object encoding/decoding using an embedded EC module, (2) load balancing the requests across a distributed set of proxies, and (3) determining where EC-encoded chunks are placed on a cluster of Lambda nodes.

\vspace{-12pt}
\paragraph{Erasure Coding Processing.}
In our initial design, we observed that adding EC processing to the proxy would stall the chunk streaming pipeline (\cref{subsec:proxy}) and significantly impact the overall data transfer performance. Hence we made a design choice to move the computation-heavy EC part from the proxy to the client library.

\vspace{-12pt}
\paragraph{The {\texttt{PUT}} Path.}
Assume that we have a multi-proxy deployment in which each proxy manages a separate Lambda node pool with shared access among clients. For a {\small\texttt{PUT}} request, {\proj}'s client library first determines the destination proxy (and therefore its backing Lambda  pool) by using a consistent hashing-based load balancing approach. The client library then encodes the object with a pre-configured EC code ($(d+p)$ using a Reed-Solomon (RS) code) and produces a number of object chunks, each with a unique identifier $ID_{obj\_chunk}$ (computed as a concatenation of the object key and the chunk's sequence number). To handle extremely large objects, {\proj} can encode them with more aggressive EC code (e.g., $(20+4)$).
Next, the client decides which Lambda nodes to store the chunks on by randomly generating a vector of non-repetitive $ID_{\lambda}$.
Each encoded chunk with its piggybacked <$ID_{obj\_chunk}, ID_{\lambda}$> is sent to the destination proxy, which streams the data to the destination Lambda nodes and remembers the locations in the Lambda pool where the chunks are cached.

\vspace{-12pt}
\paragraph{The {\texttt{GET}} Path.}
A {\small\texttt{GET}} request is first sent to the proxy by using consistent hashing; the proxy then consults its mapping table, which records the chunk to Lambda node association and fetches the object chunks from the associated Lambda nodes (see \cref{subsec:proxy}). Once the chunks arrive at the client, the client library decodes the chunks, reconstructs the original object, and returns the object to the application. 

\begin{figure}[t]
\begin{center}
\vspace{-5pt}
\includegraphics[width=0.35\textwidth]{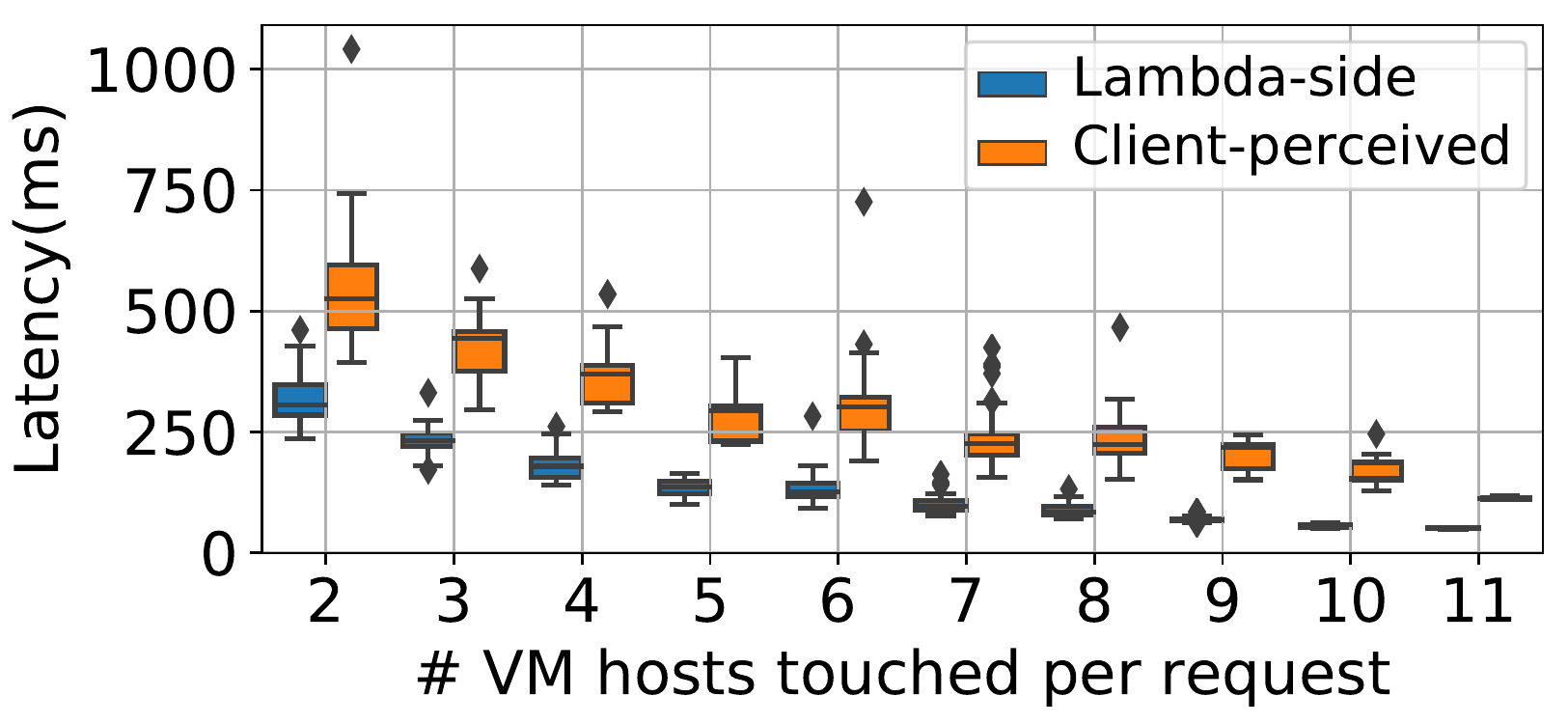}
\vspace{-10pt}
\caption{
The box-and-whisker plot of latencies as a function of the number of VM hosts touched per request.
}
\vspace{-25pt}
\label{fig:lambda_contetion}
\end{center}
\end{figure}

\vspace{-12pt}
\paragraph{Eliminating Lambda Contention.}
Lambda functions are hosted by EC2 Virtual Machines (VMs). A single VM can host one or more functions.
AWS seems to provision Lambda functions on the smallest possible number of VMs using a greedy binpacking heuristic~\cite{peeking_atc18}. This could cause severe network bandwidth contention if multiple network-intensive Lambda functions get allocated on the same host VM.

We conduct an empirical study to verify this. In our study setup, each Lambda function has 256 MB memory. We use an RS code of $(10+1)$ to split a 100 MB object into $10$ data chunks and $1$ parity chunk, and place each chunk on a Lambda node randomly selected from a fixed sized Lambda node pool. We measure the latency of {\small\texttt{GET}} requests by scaling-up the pool from 20 to 200 Lambda nodes. As a result, the number of host VMs that the 11-chunk object spans varies proportionally as the Lambda node pool scales up and down\footnote{We run command {\small\texttt{uname}} in Lambda to get the underlying host VM's IP.}. Figure~\ref{fig:lambda_contetion} shows the latency distribution as a function of the number of underlying host VM touched per request.
With a larger Lambda node pool (where the request is more likely to be spread across more host VMs), we observe a decreasing trend in the latency on the Lambda-side (the time that each Lambda node spends serving the chunk request) as well as the client-perceived (end-to-end) latencies.

These results stress the need to minimize resource contention among multiple Lambda functions sharing the same VM host. While over-provisioning a large Lambda node pool with many small Lambda functions would help to statistically reduce the chances of Lambda co-location,
we find that using relatively bigger Lambda functions largely eliminates Lambda co-location. 
Lambda's VM hosts have approximately 3 GB memory. As such, if we use Lambda functions with $\geq$ 1.5 GB memory, every VM host is occupied exclusively by a single Lambda function, assuming {\proj}'s cache pool consists of Lambda functions with the same configuration\footnote{AWS does not allow sharing Lambda-hosting VMs across tenants~\cite{lambda_whitepaper}.}.

\vspace{-12pt}
\subsection{Proxy}
\label{subsec:proxy}
\vspace{-6pt}

\begin{figure}[h]
\begin{center}
\includegraphics[width=0.4\textwidth]{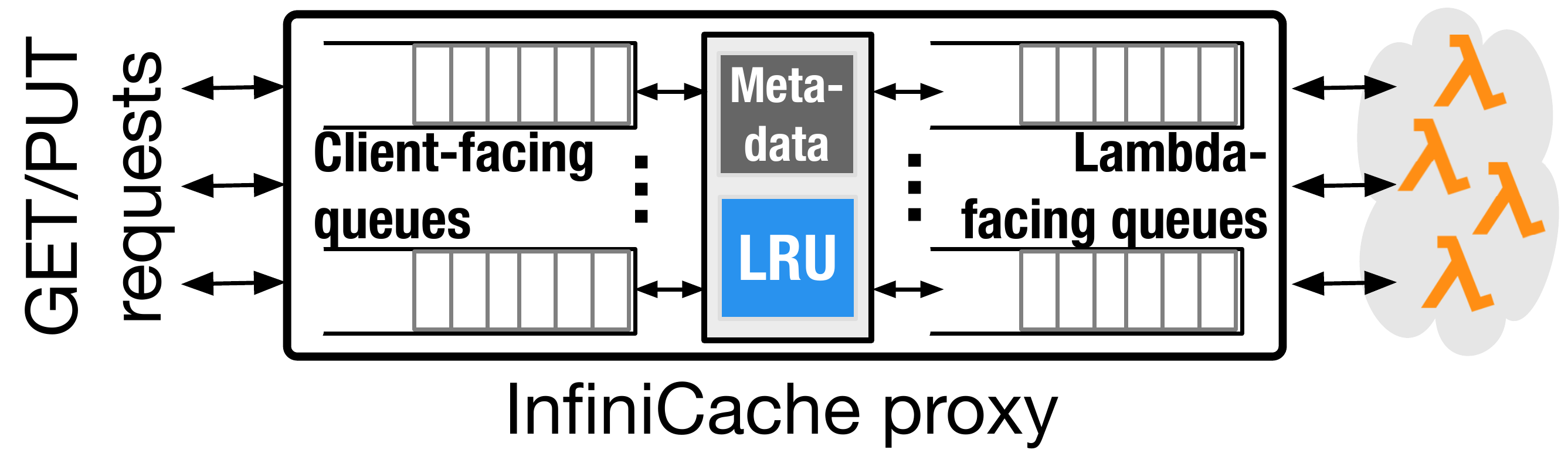}
\vspace{-7pt}
\caption{{\proj} proxy.}
\label{fig:proxy}
\vspace{-25pt}
\end{center}
\end{figure}

Each {\proj} proxy (Figure~\ref{fig:proxy}) is responsible for: (1) managing a pool of Lambda nodes, and (2) streaming data between clients and the Lambda nodes. 
Each Lambda node proactively establishes a persistent TCP connection with its managing proxy.

\vspace{-12pt}
\paragraph{Pool Management.}
Each proxy manages a pool of Lambda nodes,
and also maintains the metadata to record the mapping between object chunks and Lambda nodes. 
To achieve fault tolerance, the proxy also serves as a coordinator to coordinate data migration and delta sync (see detail in \cref{subsec:ft}).
Each proxy tracks the memory usage of every Lambda node in the pool. The proxy starts to evict objects 
as long as there is not enough free memory in the Lambda pool using a CLOCK based~\cite{clock_lru} LRU policy. The LRU module operates at the object granularity at the proxy. After the eviction process, the proxy updates the mapping metadata, and inserts the new data.

\vspace{-12pt}
\paragraph{First-d based Parallel I/O.}
The proxy sends and receives object chunks in parallel by utilizing I/O parallelism to maximize network bandwidth utilization. To mitigate the Lambda straggler problem, the proxy directly streams the first $d$ out of $(d+p)$ encoded object chunks to the client.
Though accepting the first-d arrived chunks may likely result in an EC decoding process at the client library, as we show in \cref{subsec:microbenchmark}, the performance benefit of the optimization outweights the EC decoding overhead with reduced tail latency for {\small\texttt{GET}} requests.

\vspace{-10pt}
\subsection{Lambda Function Runtime}
\label{subsec:lambda}
\vspace{-6pt}

The Lambda function runtime executes inside each Lambda instance and is designed to manage the cached object chunks in the function's memory. Our Lambda runtime uses several techniques to work around the inherent limitations of AWS Lambda. These techniques, as described below, ensure that caching is robust and cost-effective with negligible overhead.

\vspace{-12pt}
\paragraph{Memory and Connection Management.}
The Lambda runtime tracks cached key-value pairs that are sorted with a CLOCK-based priority queue\footnote{Note that CLOCK is being leveraged for two unrelated purposes: per-proxy for object eviction (\cref{subsec:proxy}), and per-node for chunk backup ordering.} for facilitating the ordered chunk backup process described in \cref{sec:designing-for-fault-tolerance}. 
Since AWS Lambda does not allow inbound TCP or UDP connections, each Lambda runtime establishes a TCP connection with its designated proxy server, the first time it is invoked. A Lambda node gets its proxy's connection information via its invocation parameters.
The Lambda runtime then keeps the TCP connection established until reclaimed by the provider.

\vspace{-12pt}
\paragraph{Anticipatory Billed Duration Control.}
AWS charges Lambda usage per 100 ms (which we call a \emph{billing cycle}). To maximize the use of each billing cycle and to avoid the overhead of restarting Lambdas, {\proj}'s Lambda runtime uses a timeout scheme to control how long a Lambda function runs.
When a Lambda node is invoked by a chunk request, a timer is triggered to limit the function's execution time. The timeout is initially set to expire within the first billing cycle. The runtime employs a simple heuristic to decide whether to extend the timeout window.
If no further chunk request arrives within the first billing cycle, the timer expires and returns 2--10 ms (a short time buffer) before the 100 ms window ends. This avoids accidentally executing into the next billing cycle.
The buffer time is configurable, and is empirically decided based on the Lambda function's memory capacity. If more than one request can be served within the current billing cycle, the heuristic extends the timeout by one more billing cycle, anticipating more incoming requests.

\begin{figure*}[t]
\begin{minipage}{\textwidth}
\begin{minipage}{0.5\textwidth}
  \centering
  \includegraphics[width=0.85\textwidth]{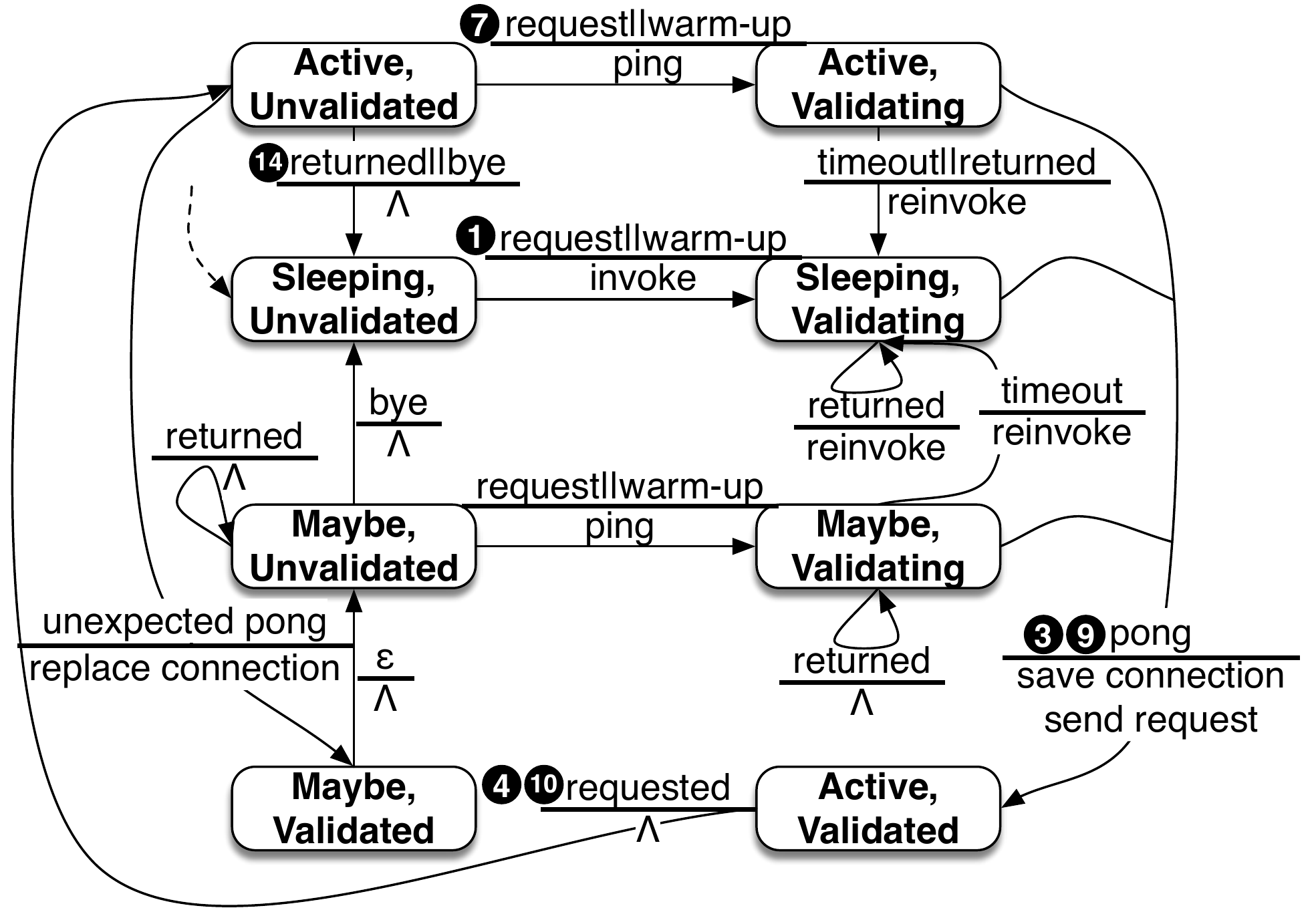}
\vspace{-10pt}
    \caption{Lambda connection validation process in a proxy.}
    \label{fig:proxy_connection_state}
\end{minipage}
\hfill
\begin{minipage}{0.5\textwidth}
  \centering
  \includegraphics[width=0.85\textwidth]{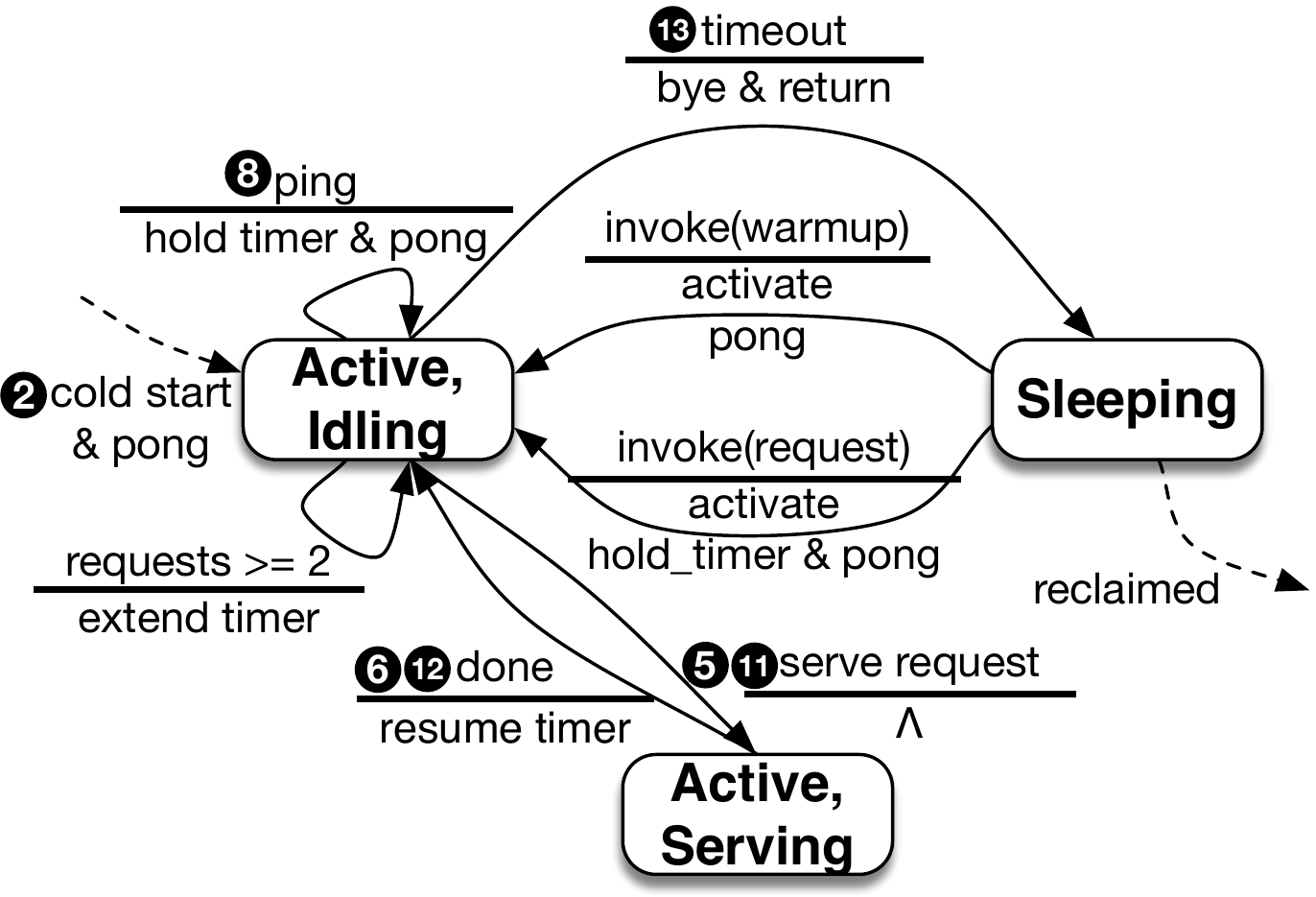}
\vspace{-10pt}
  \caption{State transitions of a Lambda function runtime.}
  \label{fig:lambda_state}
\end{minipage}
\end{minipage}
\vspace{-20pt}
\end{figure*}

\vspace{-12pt}
\paragraph{Preflight Message.}
While the proxy knows whether a Lambda node is running or has already returned, it does not know when a Lambda node will expire and return. Because of the billed duration control design that was just described, a Lambda node may return at any time. For example, right after the proxy has sent a request but before the request arrives at the Lambda, the Lambda function may expire, resulting in a denial of the request.
The proxy could maintain global knowledge about the Lambda node's real-time states by periodically polling the Lambda node. However, this is costly especially if the Lambda pool size scales up to several thousand nodes.

To eliminate such overhead, the proxy issues a preflight message ({\small\texttt{PING}}) each time a chunk request is forwarded to the Lambda node. Upon receiving the preflight message, the Lambda runtime responds with a {\small\texttt{PONG}} message,  delays the timeout (by extending the timer long enough to serve the incoming request), and when the request has been served, adjusts the timer to align it with the ending of the current billing cycle.
To further reduce overhead, the proxy can attach the {\small\texttt{PING}} message as a parameter of a Lambda function invocation request,
if the Lambda node is in sleep mode (i.e., not running but cached by AWS). Once awoken, the Lambda runtime sends a {\small\texttt{PONG}} response back to the proxy.

\vspace{-10pt}
\subsection{Reliable Lambda Connections}
\label{subsec:reliable_connection}
\vspace{-6pt}

To maintain reliable network connections between Lambda nodes and their proxy, each proxy lazily validates the status of a Lambda node every time there is a request to send. A proxy maintains three states for each Lambda connection:
1)~A {\small\texttt{Sleeping}} state---a Lambda node that is not actively running;
2)~An {\small\texttt{Active}} state---an actively running Lambda node;
3)~A {\small\texttt{Maybe}} state---during data backup~(\cref{sec:designing-for-fault-tolerance}) the original Lambda connection might have been temporarily replaced with a new connection connecting the proxy to the destination Lambda node. Figure~\ref{fig:proxy_connection_state} and Figure~\ref{fig:lambda_state} depict the state transition graphs for the proxy and  the Lambda function runtime, respectively. Note the step numbers show the interactions between a proxy (Figure~\ref{fig:proxy_connection_state}) and a Lambda function (Figure~\ref{fig:lambda_state}).

\vspace{-12pt}
\paragraph{Connection Lifecycle.}
Initially, no Lambda node is connected to the proxy. The connection is {\small\texttt{(Sleeping,Unvalidated)}}. \circled{1} When a request comes, or if a pre-warm-up is necessary, \circled{2} the proxy invokes a Lambda node. 
\circled{3} Once the Lambda node is actively running and has successfully connected to its proxy, the Lambda runtime sends a {\small\texttt{PONG}} message to proxy. Now the connection's state becomes {\small\texttt{(Active,Validated)}}, and the proxy can start issuing chunk requests. \circled{4} After the proxy sends a chunk request, the connection transits to the state {\small\texttt{(Active,Unvalidated)}}.
Having served the request (\circled{5} transits from {\small\texttt{Active,Idling}} to {\small\texttt{Active,Serving}} while \circled{6} transits back), if the proxy forwards the next request continuously, a re-validation of the connection is necessary. \circled{7} A {\small\texttt{PING}} message is sent. \circled{8} This time, the Lambda node replies with a {\small\texttt{PONG}} directly, which \circled{9} makes the connection {\small\texttt{(Active,Validated)}} again, and \circled{10} the proxy continues to issue the next chunk request. Note that the Lambda node may return anytime, or a message may timeout. In this case, the proxy re-invokes the Lambda node while marking the connection as {\small\texttt{(Sleeping,Validating)}}. Having served the request (\circled{11} transits from {\small\texttt{Active,Idling}} to {\small\texttt{Active,Serving}} while \circled{12} transits back), if no request arrives, the Lambda node \circled{13} sends {\small\texttt{BYE}} to the proxy and returns, and then the proxy \circled{14} transits the connection state back to {\small\texttt{(Sleeping,Unvalidated)}}.

When a connection is in the {\small\texttt{Maybe}} state, it behaves like an {\small\texttt{Active}} connection except that the proxy ignores the ``return'' of the source Lambda node. This does not cause a correctness issue since the source has already been replaced by a new one (i.e., the destination). The connection is marked as {\small\texttt{(Sleeping,Unvalidated)}} if a {\small\texttt{BYE}} message is received via the connection.

\vspace{-12pt}
\section{Data Availability and Fault Tolerance}
\label{subsec:ft}
\vspace{-7pt}

In this section, we conduct a case study with AWS Lambda and describe the approaches {\proj} employs for maintaining practical data availability and fault tolerance over a fleet of ephemeral cloud functions with a high churn rate.

\vspace{-10pt}
\subsection{AWS Lambda Properties}
\label{subsec:aws-lambda-properties}
\vspace{-4pt}

While AWS allows function caching to mitigate ``cold start'' overhead, it does not provide any availability guarantees for the cached function and can reclaim it anytime. Hence, {\proj} needs to be robust against frequent failures of cache nodes.
To better understand the stateless property of AWS Lambda and its implications on short-term data availability, we conduct an extensive black-box analysis. We analyze reclamation behaviors by quantifying the number of reclaimed Lambda functions in a 24-hour period under different warm-up strategies.

According to a recent study~\cite{peeking_atc18}, a Lambda function that finishes execution is kept by AWS for at most 27 minutes if that function is not invoked again.
A function's lifespan can be extended to hours if that function instance is invoked periodically (i.e., by so-called warm-up operations). The lifespan extension varies according to the warm-up strategy as well as AWS' internal resource management policy.

We deploy a pool of 300--400 Lambda functions with the same memory configuration, and re-invoke each one of them every $N$ minute(s). Each function simply returns an ID value that the function computed when it was invoked the first time.
If AWS reclaims an already invoked, cached function, a new function instance will be instantiated at the next invocation request and the ID of this function will change. We keep track of the ID to detect whether a function has been reclaimed or not. We evaluate two warm-up strategies: a low warm-up frequency (every 9 minutes) and a relatively high frequency (every 1 minute). We ran each strategy during a span of 6 months (from August 2019 to January 2020), and recorded the number of function reclaiming events.

\begin{figure}[t]
\begin{center}
\vspace{-5pt}
\includegraphics[width=0.35\textwidth]{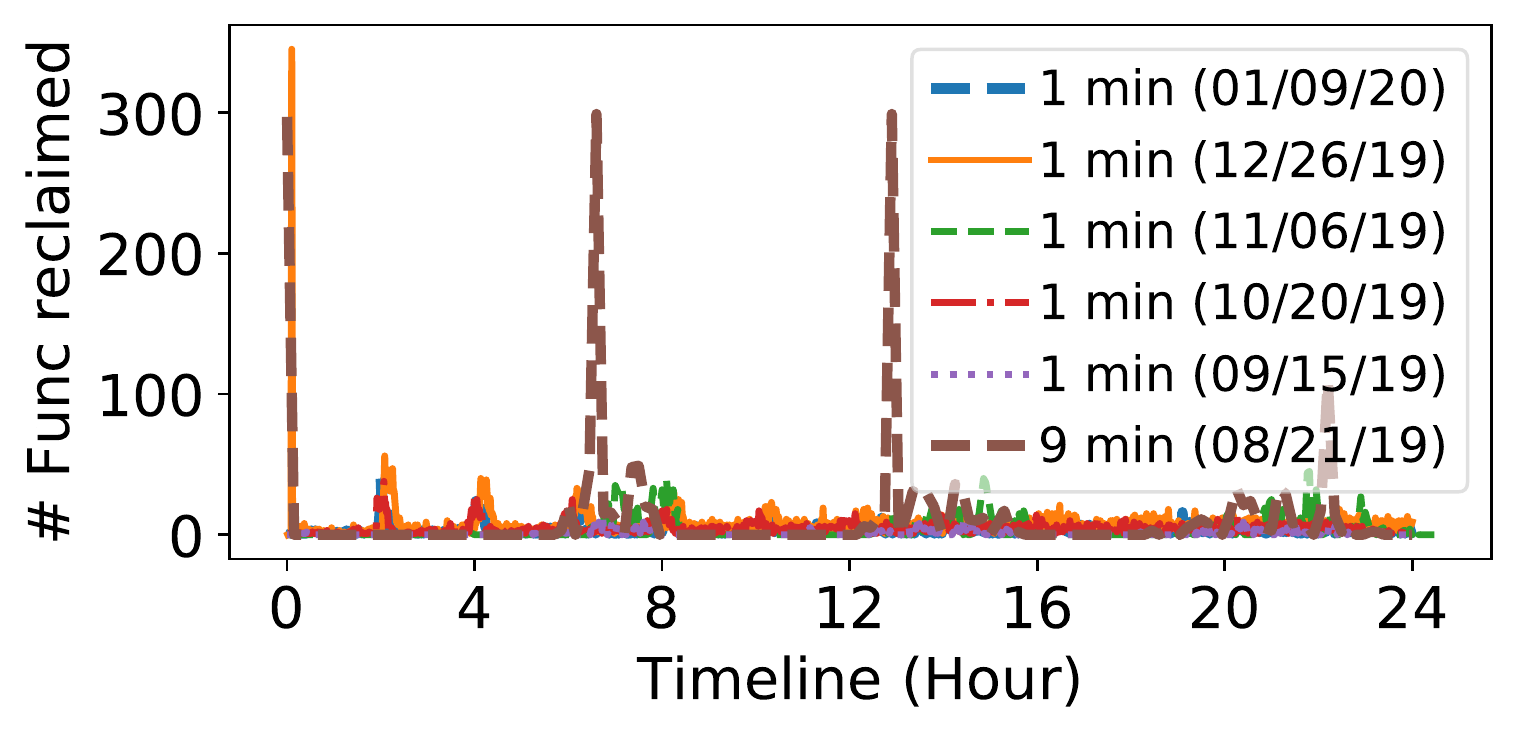}
\vspace{-10pt}
\caption{Number of functions being reclaimed over time under various warm-up strategies. 
}
\vspace{-18pt}
\label{fig:reclaim_timeline}
\end{center}
\end{figure}

As shown in Figure~\ref{fig:reclaim_timeline}, for {\small\texttt{9 min (08/21/19)}}, we observe a large number of function reclaiming events clustered around hour~6, hour~12, and hour~20--22. The number of reclaimed functions spiked roughly every 6 hours and almost all the functions get reclaimed. For {\small\texttt{1 min (09/15/19)}}, the situation got much better; the peak number of reclaiming events gets reduced to 22, 21, and 16 at hour~6, respectively. Similar trends appeared in November, but got substantially changed in December and January -- for example for {\small\texttt{1 min (12/26/19)}},
instead of spiking every 6 hours, AWS continuously reclaimed Lambda functions with an hourly reclaiming rate of 36. This is possibly due to AWS Lambda's internal policy changes after AWS announced the launch of provisioned concurrency~\cite{pconcurrency} for Lambda on December 03, 2019.

\begin{figure}[t]
\begin{center}
\vspace{-5pt}
\includegraphics[width=0.36\textwidth]{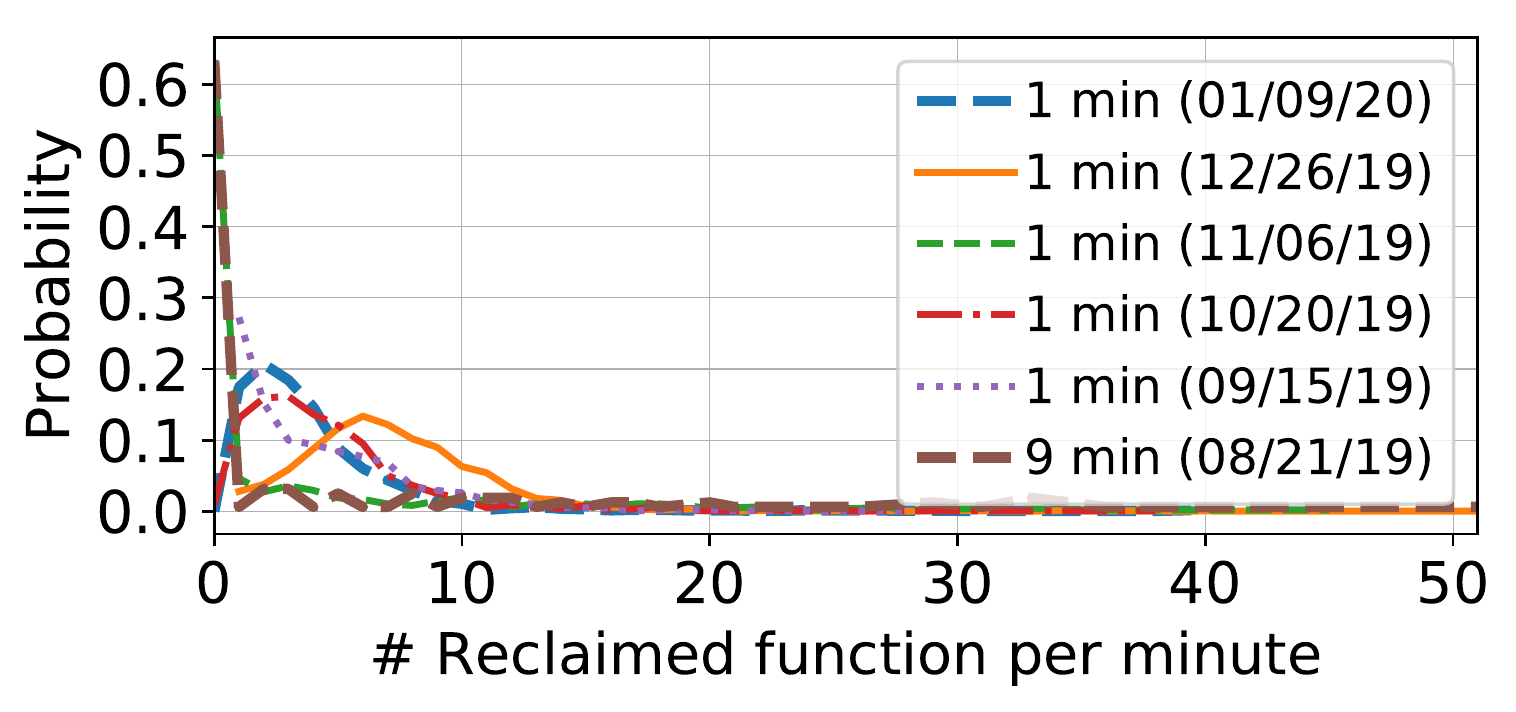}
\vspace{-10pt}
\caption{Probability distribution of the number of functions reclaimed per minute on the sampled days. }
\vspace{-20pt}
\label{fig:reclaim_distribution}
\end{center}
\end{figure}

Figure~\ref{fig:reclaim_distribution} shows the function reclaiming events roughly follow a Zipf distribution for August, September, and November (with different $s$ values), and a Poisson distribution for October, December, and January (with different $\lambda$ values). With that, we can calculate an approximate range of probabilities of $r$ functions being reclaimed simultaneously in a user-defined interval (\cref{sec:data-availability-analysis}). 
Motivated by these observations, we argue that with careful design, we can improve the data availability for {\proj}.

\vspace{-10pt}
\subsection{Maximizing Data Availability}
\label{sec:designing-for-fault-tolerance}
\vspace{-5pt}

{\proj} adopts three techniques for maximizing data availability: (1)~EC is used to enable data recovery for up to $p$ object chunk losses given an RS code $(d+p)$. In the case that there are more than $p$ chunks lost, tenants need to retrieve the data from the backing object store; (2)~each Lambda runtime is warmed up after every $T_{warm}$ interval of time; we use a $T_{warm}$ value of of 1 minute as motivated by our observations in \cref{subsec:aws-lambda-properties}; (3)~to further enhance availability, a delta-sync based data backup scheme provides incremental backups every  $T_{bak}$ interval. The selection of $T_{bak}$ is a trade-off between availability, runtime overhead, and cost effectiveness: a shorter interval lowers the data loss rate while a longer interval incurs less backup overhead and less cost. In the following, we explain the backup scheme in more detail.

\begin{figure}[t]
\begin{center}
\includegraphics[width=0.4\textwidth]{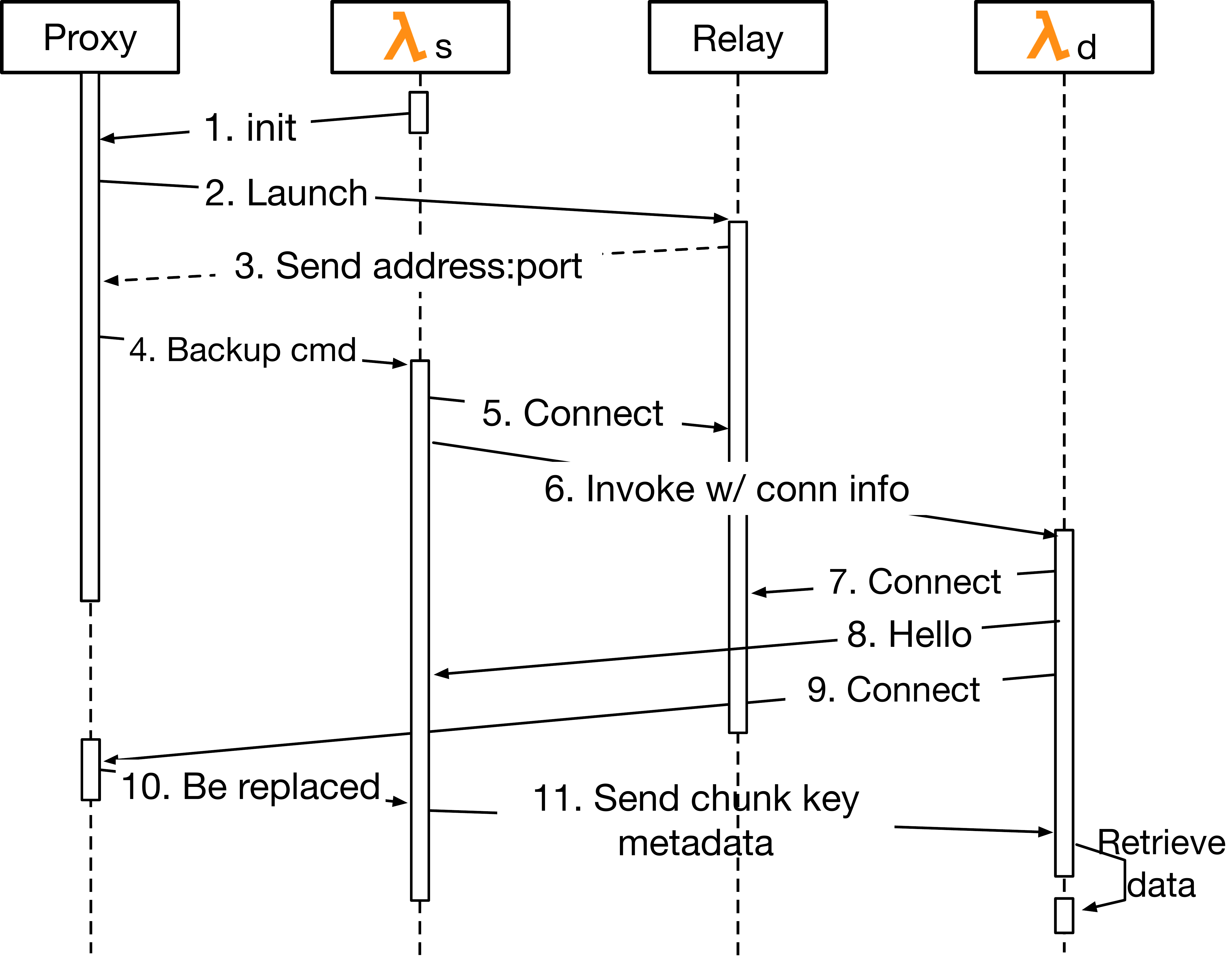}
\vspace{-10pt}
\caption{{\proj}'s backup protocol. 
}
\vspace{-30pt}
\label{fig:backup_protocol}
\end{center}
\end{figure}

\vspace{-12pt}
\paragraph{Backup Protocol.}
{\proj} performs periodic delta-sync backups between two peer replicas\footnote{Concurrent invocations to the same function produce multiple concurrent Lambda instances -- this process is called auto-scaling. Here we call each instance of the same function a \emph{peer replica}.
} of the same Lambda function. We choose peer replicas instead of distinct Lambdas to be able to seamlessly failover to one of them in case the other gets reclaimed.

In light of the observations in Figure~\ref{fig:reclaim_timeline}, we design a backup scheme that preserves the following properties: 
(1)~autonomicity---a Lambda node should backup itself with minimum help from the proxy to keep proxy logic simple;
(2)~high availability---the service provided by the Lambda node should not be interrupted; and
(3)~low network overhead---large object workloads are network bandwidth sensitive so backups should cause low or no extra network overhead. To this end, we adopt an efficient, Lambda-aware mechanism that performs delta-sync between two peer replicas of the same Lambda function.

The protocol sequence graph is depicted in Figure~\ref{fig:backup_protocol}.
In Step 1, a Lambda node $\lambda_s$, serving as the source cache node, sends an {\small\texttt{init-backup}} message to its proxy to initialize a backup process every $T_{bak}$. Acknowledging this message, in Step 2, the proxy launches a new process called \emph{relay} (co-allocated with proxy), which serves to forward TCP packets between $\lambda_s$ and a destination Lambda node $\lambda_d$, i.e., the Lambda node that receives the backup. In Step 3, the relay process sends its own network information (address:port) to the proxy, which issues a {\small\texttt{backup}} command in Step 4 to $\lambda_s$, piggybacked with the relay's connection information.

In Step 5, $\lambda_s$ establishes a TCP connection with the relay and in Step 6 invokes a peer replica instance of the $\lambda_s$ function, which serves as $\lambda_d$; at the same time, $\lambda_s$ passes the connection information of both the relay and proxy to $\lambda_d$ as the Lambda invocation parameters. In Step 7, $\lambda_d$ establishes a TCP connection with the relay. If connected successfully, an indirect network channel is bridged through a relay between $\lambda_s$ and $\lambda_d$. Then $\lambda_d$ sends a {\small\texttt{hello}} message to $\lambda_s$ in Step 8 and connects to the proxy in Step 9.

Upon establishing the connection with $\lambda_d$, in Step 10, the proxy disconnects from $\lambda_s$, which makes $\lambda_d$ the only active connection to the data of $\lambda_s$. Hence, the proxy forwards all requests to $\lambda_d$ while $\lambda_d$ forwards requests to $\lambda_s$, if it has not yet received the requested data.
To receive data, $\lambda_d$ sends a {\small\texttt{hello}} to $\lambda_s$ in Step 11 and $\lambda_s$ starts sending metadata (stored chunk keys) in an order from MRU to LRU.
Once $\lambda_d$ has received all the keys, it starts the data migration by retrieving the data associated with the keys from $\lambda_s$.

If $\lambda_d$ receives a {\small\texttt{PUT}} request during data retrieval and the key is not found, it inserts the new data in its cache and then forwards it to $\lambda_s$.
If a {\small\texttt{GET}} request is received for a key that has been retrieved already from $\lambda_s$, $\lambda_d$ directly responds with the requested chunk. Otherwise, $\lambda_d$ forwards the request to $\lambda_s$, responds to the proxy, and then caches the key and the corresponding chunk.

After data retrieval completes, $\lambda_d$ returns and the connection to the proxy becomes inactive. Hence, the next time the proxy invokes this Lambda function, AWS would launch one of the two, $\lambda_s$
or $\lambda_d$, if they have not been reclaimed yet. As they are now in sync, they can both serve the data.
After another interval $T_{bak}$, the whole backup procedure repeats. $\lambda_d$ only retrieves the ``delta'' part of data to reduce overhead.

\vspace{-10pt}
\subsection{Data Availability and Cost Analysis}
\label{sec:data-availability-analysis}
\vspace{-6pt}

\paragraph{Availability Analysis.}
To better understand the data availability of {\proj}, we build an analytical model. 
Assume $N_\lambda$ is the total number of Lambda nodes. At time $T_r$, a number $r$ of nodes are found reclaimed. 
$m$ is the minimum number of chunks that leads to an object loss and $n$ is the number of EC chunks of a object. An object is considered not available if there are at least $m$ chunks lost due to function reclaiming. The probability $P(r)$ that an object is not available (i.e., lost) is formalized as: $P(r) = \sum_{i=m}^{n}p_i$, where:

\vspace{-5pt}
\begin{equation}
\small
    p_i = \frac{C(r, i)C(N_\lambda - r, n-i)}{C(N_\lambda, n)}.
\vspace{-4pt}
\end{equation}

\noindent  Here $C(r, i)$ is the combinations in which $r$ reclaimed Lambda nodes happens to hold $i$ chunks belonging to the same object. 
$C(N_\lambda - r, n-i)$ is the combinations in which the rest chunks of that object are held in Lambda nodes that have not been reclaimed.
$C(N_\lambda, n)$ is the combinations in which all Lambda nodes hold all chunks of an object.

Assuming $p_d(r)$ is the probability distribution of reclaiming $r$ Lambda nodes at $T_r$, the probability of losing an object $P_l$ is the sum of the probabilities of losing one object when at least $m$ Lambda nodes are reclaimed:

\vspace{-12pt}
\begin{equation}
\small
\begin{aligned}
    P_l = \sum_{r=m}^{N_\lambda}P(r)p_d(r) = \sum_{r=m}^{N_\lambda}\sum_{i=m}^{n}\frac{C(r, i)C(N_\lambda - r, n-i)}{C(N_\lambda, n)}p_d(r).
\end{aligned}
\vspace{-5pt}
\end{equation}

One observation is that $\frac{p_m}{p_{m+1}}$ can be larger than 10.
E.g., for a 400-Lambda nodes deployment with $N_\lambda = 400$, an RS code of $(10+2)$, and a warm-up interval of 1 minute, if 12 nodes get reclaimed simultaneously at time $T_r$, we have $p_3/p_4 = 18.8$ for $r=12$, and $P(r)$ is only about $5\%$ larger than $p_3$.
So we can simplify the formulation as $P(r) \approx p_m$, thus $P_l$ can be simplified as:

\vspace{-7pt}
\begin{equation}
\small
\label{eq:p_loss}
    P_l \approx \sum_{r=m}^{N_\lambda}\frac{C(r, m)C(N_\lambda - r, n-m)}{C(N_\lambda, n)}p_d(r).
\vspace{-2pt}
\end{equation}

In our case study, 
$N_\lambda=400$, $n=12$, $m=3$, and  $T_{warm} = 1~min$. With Equation~\ref{eq:p_loss} we get 
$P_l=0.0039\% \sim 0.11\%$ or an availability $P_a=99.89\% \sim 99.9961\%$ for 1 minute, and $93.36 \sim 99.76\%$ for 1 hour based on the variable probability distribution of Lambda reclaiming policies we observed over a six-month period (\cref{subsec:aws-lambda-properties}).

\begin{figure*}[t]
\begin{center}
\vspace{-5pt}
\subfigure[128 MB Lambda.] {
\includegraphics[width=.323\textwidth]{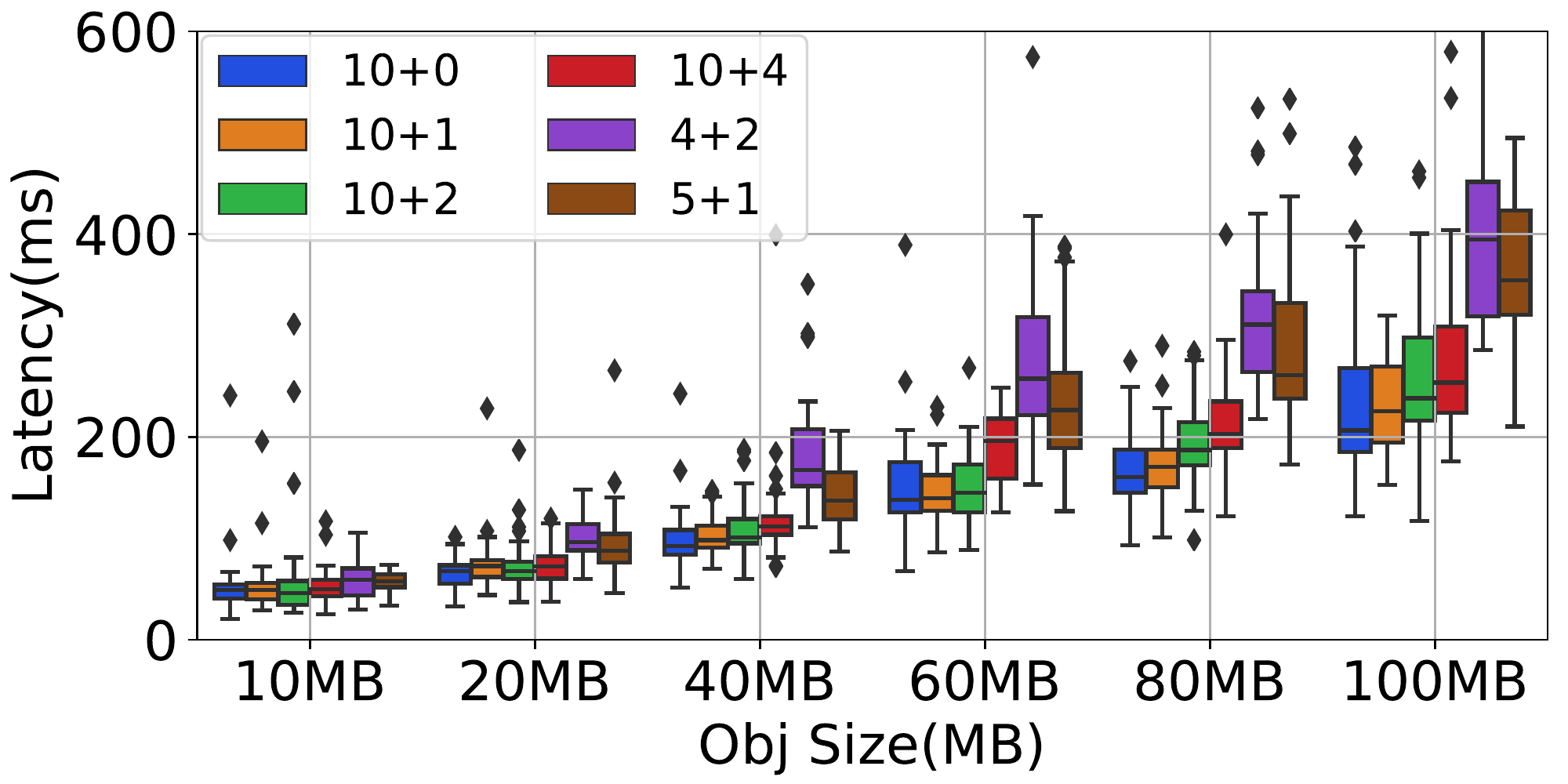}
\label{fig:micro128}
}
\hspace{-10pt}
\subfigure[256 MB Lambda.] {
\includegraphics[width=.323\textwidth]{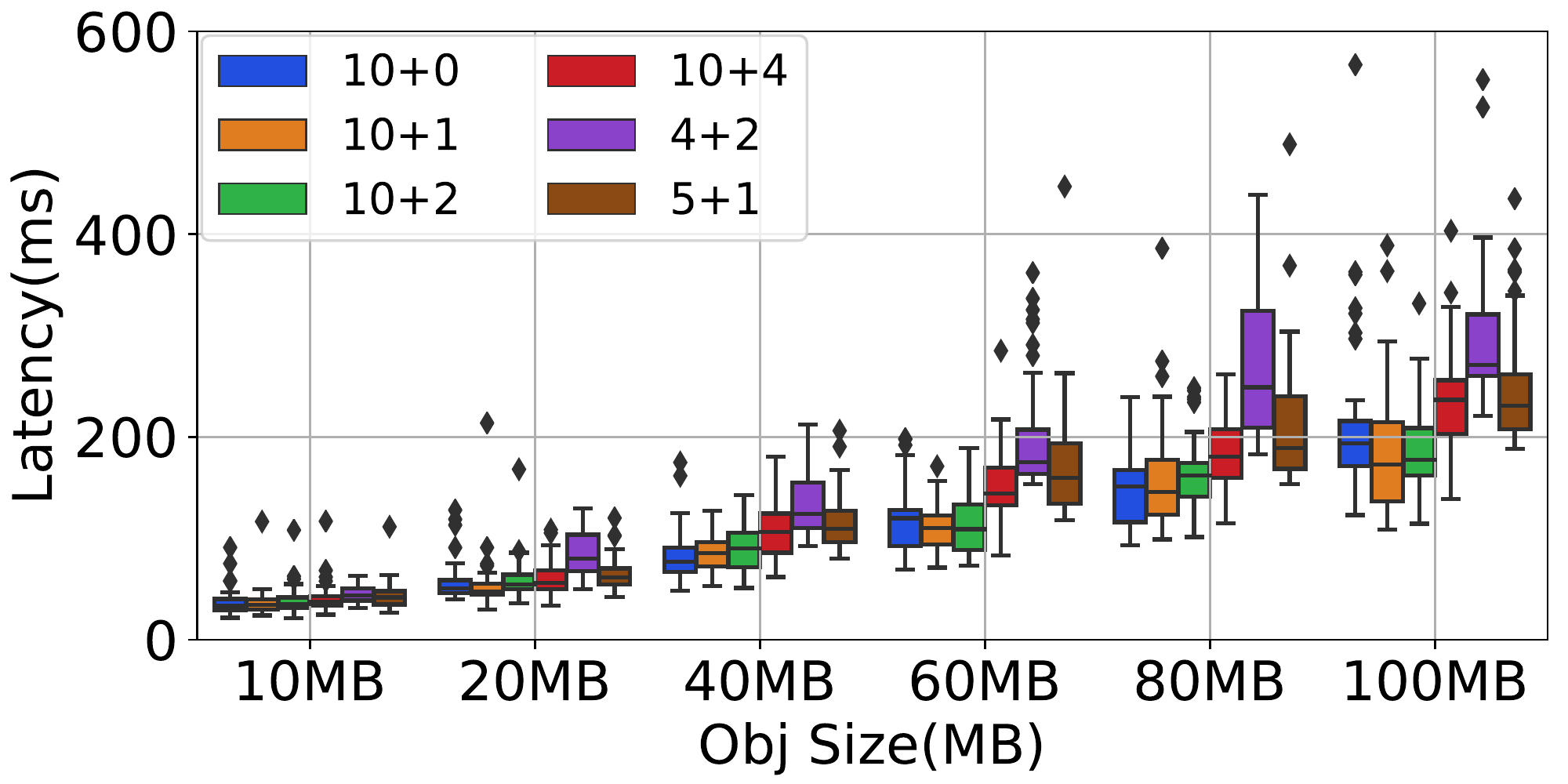}
\label{fig:micro256}
}
\hspace{-10pt}
\subfigure[512 MB Lambda.] {
\includegraphics[width=.323\textwidth]{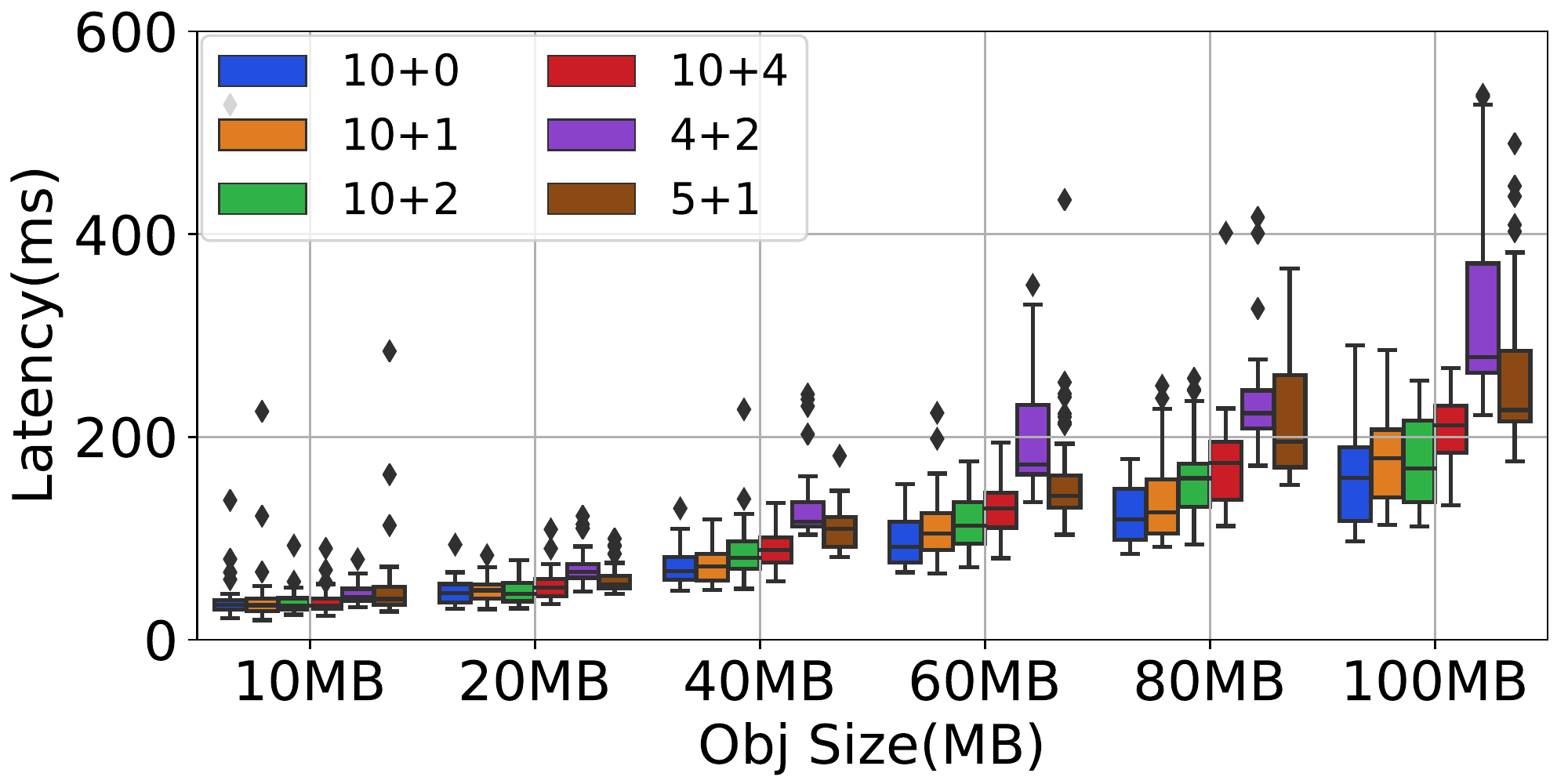}
\label{fig:micro512}
}
\subfigure[1024 MB Lambda.] {
\includegraphics[width=.323\textwidth]{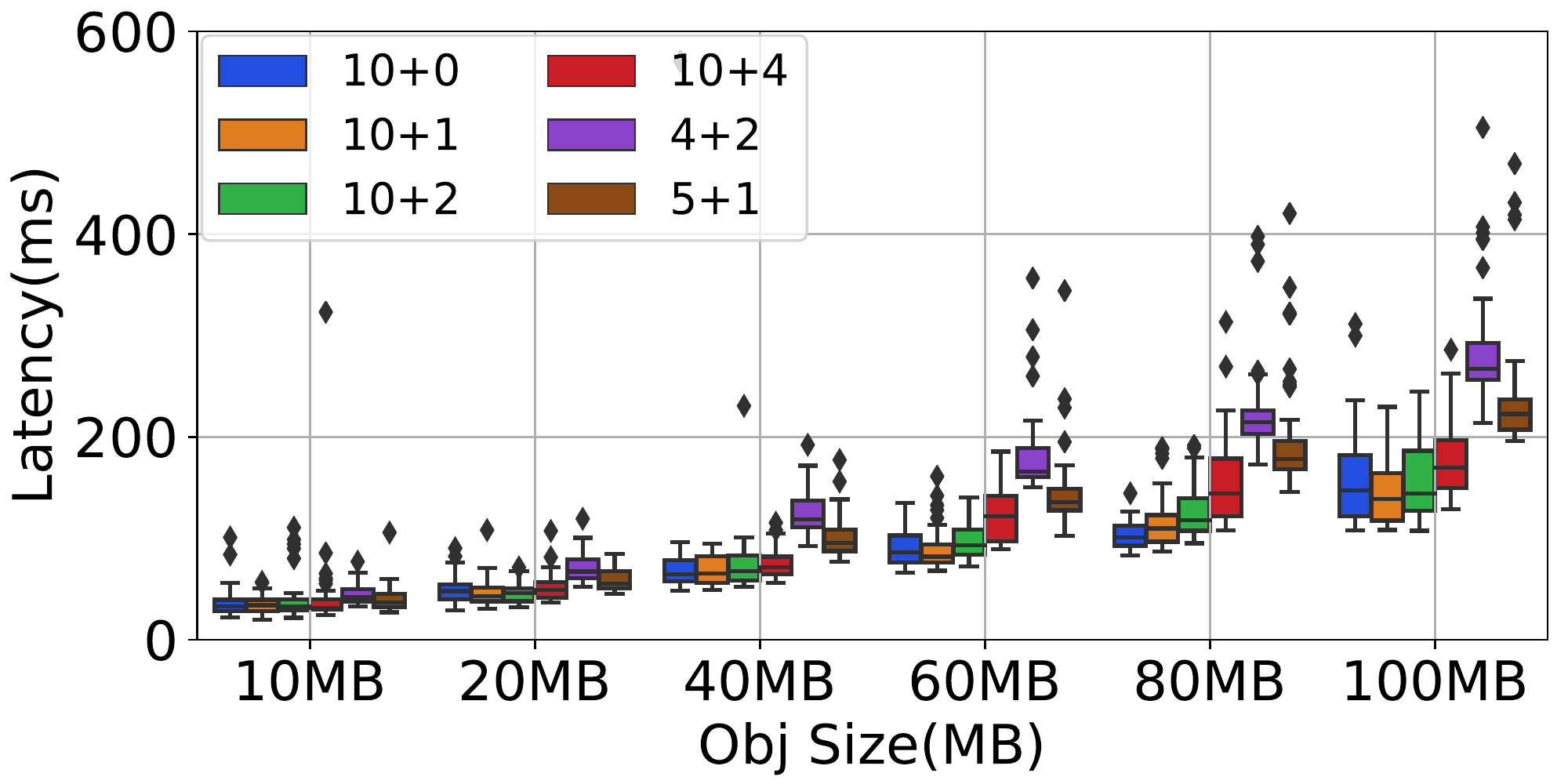}
\label{fig:micro1024}
}
\hspace{-10pt}
\subfigure[2048 MB Lambda.] {
\includegraphics[width=.323\textwidth]{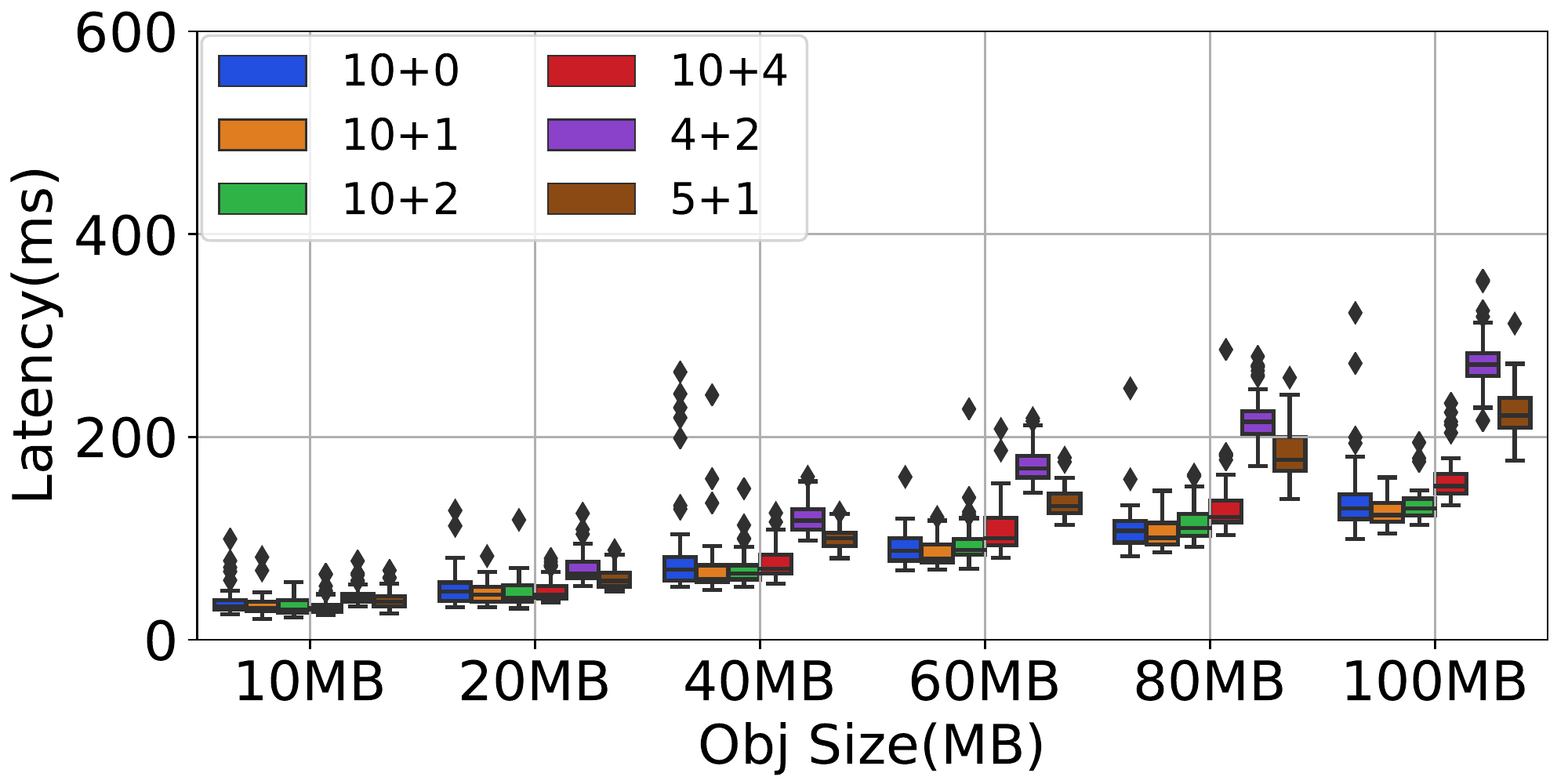}
\label{fig:micro2048}
}
\hspace{-10pt}
\subfigure[3008 MB Lambda.] {
\includegraphics[width=.323\textwidth]{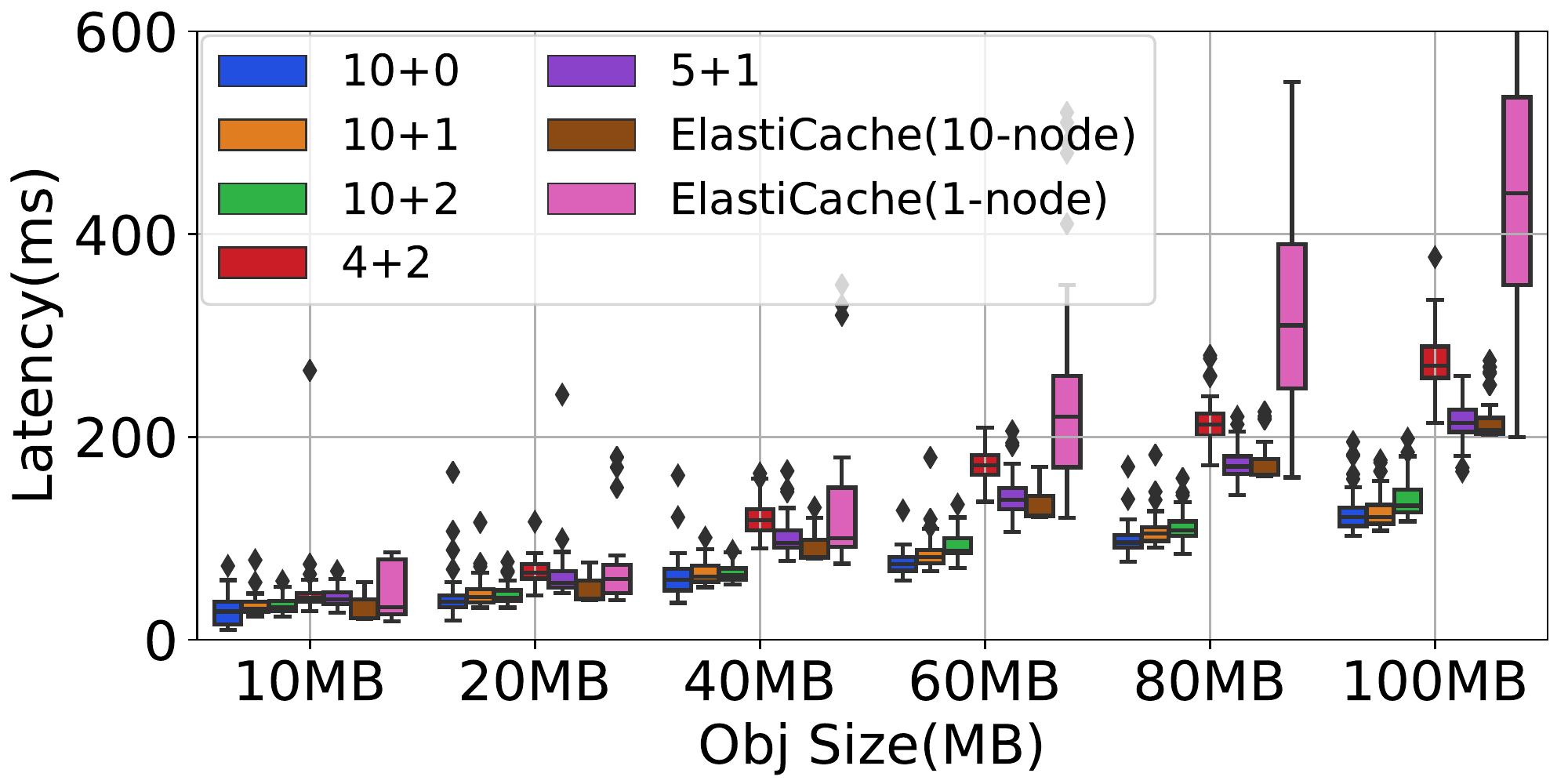}
\label{fig:micro3008}
}
\vspace{-10pt}
\caption{Microbenchmark performance. }
\label{fig:micro}
\end{center}
\vspace{-25pt}
\end{figure*}

\vspace{-12pt}
\paragraph{Cost Analysis.}
To maintain high availability, {\proj} employs EC, warm-up, and delta-sync backup, which all incur extra cost.  
For a  better understanding of how these techniques impact total cost, we build an analytical cost model.
To simplify our presentation, we do not explicitly express the EC configuration using an RS code $(d+p)$, but rather reflect it in the total number of instances $N_\lambda$. 
The total cost per hour $C$ is therefore composed of (1) serving chunk requests ($C_{ser}$), (2) warming-up  functions ($C_w$), and (3) backing up data, ($C_{bak}$). Thus, {\small $C = C_{ser} + C_w + C_{bak}$}
Next, we introduce each term respectively.

\noindent $\bullet$  {\it Serving cost $C_{ser}$.} AWS charges function invocations and function duration. We denote the price per invocation as $c_{req}$ and the duration price of per GB-second as $c_d$. The function duration is rounded up to the nearest 100 ms, we define a round-up operation $ceil_{100}(.)$. Assume Lambda's memory is $M$ GB, the average hourly request rate is $n_{ser}$, and the duration of each invocation is $t_{ser}$ ms, we have:
\vspace{-7pt}
\begin{equation}
\small
\begin{split}
C_{ser} = n_{ser}*c_{req} + n_{ser}*ceil_{100}(t_{ser})/1000*M*c_d.
\end{split}
\vspace{-5pt}
\end{equation}

\vspace{-10pt}
\noindent $\bullet$  {\it Warm-up cost $C_w$.} 
The backup frequency $f_{w} = 60/T_{warm}$.
The warm-up duration $t_w$ is typically in the range of a few ms and therefore we have $ceil_{100}(t_w) = 100 $ ms. Thus we have:
\vspace{-8pt}
\begin{equation}
\small
\begin{split}
C_w = N_\lambda*f_w*c_{req} + N_\lambda*f_w*0.1*M*c_d.
\end{split}
\vspace{-12pt}
\end{equation}

\vspace{-8pt}
\noindent $\bullet$  {\it Backup cost $C_{bak}$.} The backup frequency is denoted as $f_{bak} = 60/T_{bak}$. 
We have:
\vspace{-5pt}
\begin{equation}
\small
\begin{split}
C_{bak} = N_\lambda*f_{bak}*c_{req} + N_\lambda*f_{bak}*t_{bak}*M*c_d.
\end{split}
\vspace{-10pt}
\end{equation}

As shown in \cref{subsec:prod}, the backup cost is a dominating factor whose proportion increases as more data are being cached.
\vspace{-10pt}
\section{Evaluation}
\label{sec:eval}
\vspace{-6pt}

In this section, we evaluate {\proj} 
on AWS Lambda using microbenchmarks and a production workload from the IBM Docker registry~\cite{docker_fast18}.

\vspace{-12pt}
\paragraph{Implementation.}
We have implemented a prototype of {\proj} using $5,340$ lines of Go ($460$ LoC for the client library, $3,447$ for the proxy, and $1,433$ for the Lambda runtime). The EC module of the client library is implemented using the Golang reedsolomon lib~\cite{go_ec}, which uses Intel's AVX-512 for accelerating EC computation.

\vspace{-12pt}
\paragraph{Setup.}
Our experiments use AWS Lambda functions with various configurations. Unless otherwise specified, we deploy the client (with {\proj}'s client library) and proxy on {\small\texttt{c5n.4xlarge}} EC2 VM instances. The Lambda functions are in the same Amazon Virtual Private Cloud (VPC) as the EC2 instances and are equipped with a 10 Gbps network connection. The Lambda functions' network bandwidth increases with its memory amount; we observed a throughput of 50--160 MBps (from the smallest memory amount of 128 MB to the largest memory amount of 3008 MB) between a {\small\texttt{c5n.4xlarge}} EC2 instance and a Lambda function using {\small\texttt{iperf3}}.

\vspace{-12pt}
\subsection{Microbenchmark Performance}
\label{subsec:microbenchmark}
\vspace{-6pt}

\begin{figure}[t]
\begin{center}
\includegraphics[width=0.35\textwidth]{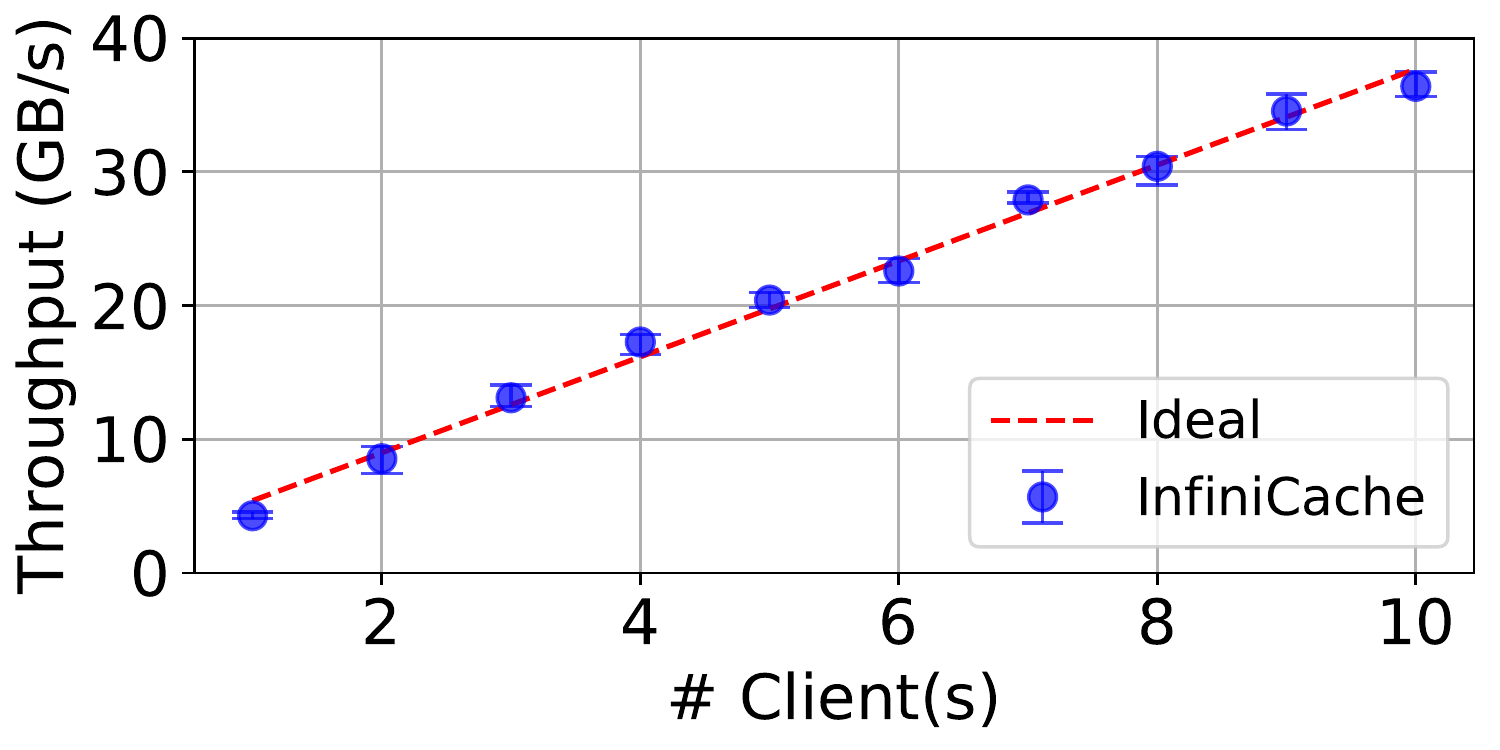}
\vspace{-10pt}
\caption{Scalability of {\proj}.}
\label{fig:scalability}
\vspace{-25pt}
\end{center}
\end{figure}

We first evaluate the performance of {\proj} under synthetic {\small\texttt{GET}}-only workloads generated using a simple benchmark tool. 
With the microbenchmarking tests, we seek to understand how different configuration knobs impact {\proj}'s performance. The evaluated configuration knobs include: EC RS code (we compare $(10+1)$, $(10+2)$, $(4+2)$, $(5+1)$,  with a $(10+0)$ baseline, which directly splits an object into 10 chunks without EC encoding/decoding), object sizes (10--100 MB), and the Lambda function's resource configurations (128--3008 MB).

Figure~\ref{fig:micro} shows the distributions of end-to-end request latencies seen under different configuration settings. Invoking a warm Lambda function takes about 13 ms on average (with the Go AWS SDK API), which is included in the end-to-end latency results.  We observe that the $(10+1)$ code performs best compared to other RS code configurations. This is due to two reasons. First, $(10+1)$ results in a maximum I/O parallelism factor of 10 (first-k parallel I/O is described in \cref{subsec:proxy}), and second, it keeps the EC decoding overhead at a minimum (the higher the number of parity chunks, the longer it takes for RS to decode). The caveat of using $(10+1)$ is that it trades off fault tolerance for better performance.

Another observation is that the $(10+0)$ case does not seem to lead to a better performance than that of $(10+1)$ and in several cases even sees higher tail latencies. This is due to the fact that $(10+0)$ 
suffers from Lambda straggler issues, which outweighs the performance gained by fully eliminating the EC decoding overhead. In contrast, $(10+1)$'s first-d approach adds redundancy and this request-level redundancy helps mitigate the impact of stragglers.

\begin{figure*}[t]
\begin{minipage}{\textwidth}
\begin{minipage}{1\textwidth}
  \centering
  \includegraphics[width=1\textwidth]{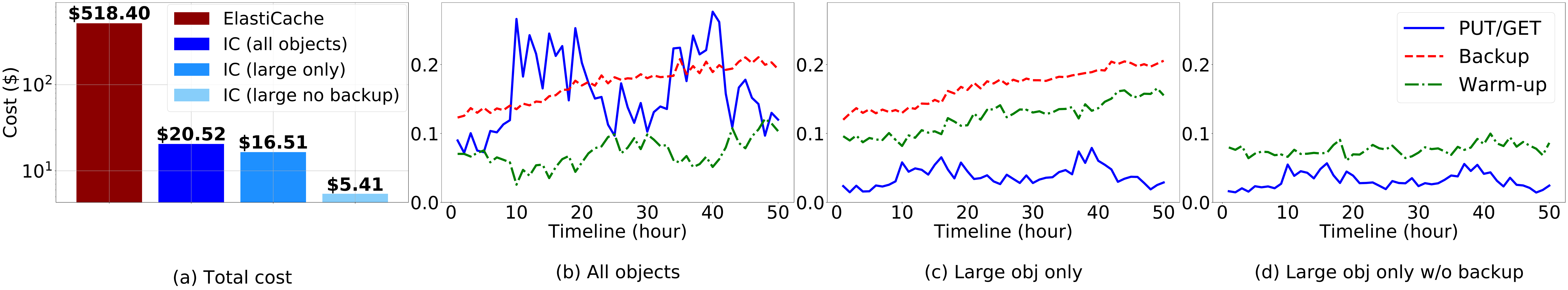}
\vspace{-20pt}
    \caption{{\small Total \$ cost (a) for ElastiCache and {\proj} ({\small\texttt{IC}}), and {\proj}'s hourly cost breakdown under various settings (b)-(e).}}
    \label{fig:docker_cost}
\end{minipage}
\hfill
\begin{minipage}{1\textwidth}
  \centering
  \includegraphics[width=1\textwidth]{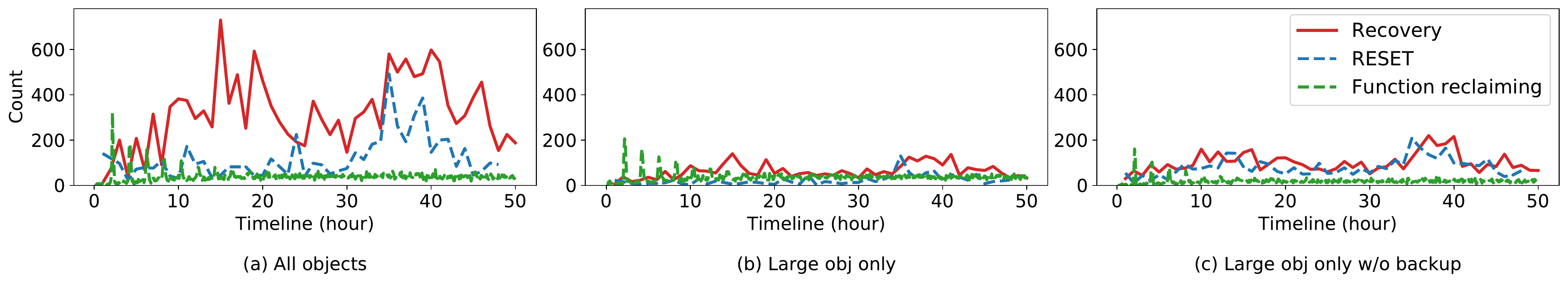}
\vspace{-20pt}
  \caption{Timeline of {\proj}'s fault tolerance activities under various workload settings.}
  \label{fig:docker_event}
\end{minipage}
\end{minipage}
\vspace{-15pt}
\end{figure*}

A Lambda function's resource configuration has a great impact on {\proj}'s latency. For example, $(10+1)$ achieves latencies in the range of 110--290 ms (Figure~\ref{fig:micro512}) with 512 MB Lambda functions for objects of 100 MB, whereas with 2048 MB Lambda functions, latencies improve to 100--160 ms (Figure~\ref{fig:micro2048}). In addition, latency improvement hits a plateau for Lambda functions equipped with more than 1024 MB memory because larger Lambda functions eliminate the network bottleneck for large chunk transfers.

To compare {\proj} with an existing solution, we choose ElastiCache (Redis) and deploy it in two modes, a {\small\texttt{1-node}} deployment using a {\small\texttt{cache.r5.8xlarge}} instance, and a scale-out {\small\texttt{10-node}} deployment using {\small\texttt{cache.r5.xlarge}} instances. As shown in Figure~\ref{fig:micro3008}, {\proj} outperforms the {\small\texttt{1-node}} ElastiCache for all object sizes, as Redis is single-threaded and cannot handle concurrent large I/Os as efficiently. For larger object sizes, {\proj} with $(10+1)$ and $(10+2)$ consistently achieves lower latencies compared to the {\small\texttt{10-node}} ElastiCache, thanks to {\proj}'s first-d based data streaming optimization. These results show that {\proj}'s performance is competitive as an IMOC.

\vspace{-12pt}
\paragraph{Scalability.}
In this test, we setup a multi-client deployment to simulate a realistic use case in which a tenant has multiple microservices that concurrently read from and write to {\proj}. To do so, we vary the number of clients from 1 to 10. We also deploy a 5-proxy cluster where each proxy manages a 50-node Lambda pool (and each Lambda function has 1024 MB memory). Each client uses consistent hashing to talk to different proxies for shared data access (see Figure~\ref{fig:arch}). 

Figure~\ref{fig:scalability} shows the throughput in terms of GB/s. We observe that {\proj}'s throughput scales linearly as the number of clients increases. Ideally, {\proj} can scale linearly as long as more Lambda nodes are available for serving {\small\texttt{GET}} requests.

\vspace{-12pt}
\subsection{Production Workload}
\label{subsec:prod}
\vspace{-7pt}

\begin{table}[t]
\begin{center}
\scalebox{0.76}{
\begin{tabular}{lrrrrr}
{\bf Workload} & {\bf WSS} & {\bf Thpt}  & {\bf EC} & {\bf IC}  & {\bf IC w/o backup}  \\ 
\Xhline{2\arrayrulewidth}
{\bf All objects} & $1,169$~GB & {3,654}  & $67.9\%$ & $64.7\%$ & -  \\
{\bf Large obj. only} & $1,036$~GB  & {750} & $65.9\%$ & $63.6\%$ & $56.1\%$  \\ 
\end{tabular}
}
\vspace{-7pt}
\caption{Workloads' working set sizes ({\small\texttt{WSS}}), throughput (average {\small\texttt{GET}}s per hour), and the cache hit ratio achieved by ElastiCache ({\small\texttt{EC}}) and {\proj} ({\small\texttt{IC}}).}
\label{tbl:wl}
\vspace{-25pt}
\end{center}
\end{table}

In this section, we evaluate {\proj} using the IBM Docker registry production workload (detailed in \cref{sec:moti}). The original workload contains a 75-day request trace spanning 7 geographically distributed datacenters. 
Out of the 7 datacenters, we select Dallas, which features the highest load. We parse the Dallas trace for {\small\texttt{GET}} requests that read a blob (i.e., a Docker image layer). We test two workload settings: 1) {\small\texttt{all objects}} (including both small and large, with a working set size (WSS) of $1,169$~GB as shown in Table~\ref{tbl:wl}), and 2) {\small\texttt{large object only}} (only including objects larger than 10~MB, with a WSS of $1,036$~GB). 

We replay the first 50 hours of the Dallas trace in \emph{real time} and skip the largest object which was 8 GB (there was only one object). A {\small\texttt{GET}} upon a miss results in a {\small\texttt{PUT}} that inserts the object into the cache. {\proj} is configured with a pool consisting of 400 1.5~GB Lambda functions, which are managed by one proxy co-located with our trace replayer as the client. We use an EC RS configuration of $(10+2)$ to balance performance with fault tolerance. We select a warm-up interval $T_{warm}$ as 1 minute (due to our study in Figure~\ref{fig:reclaim_timeline}) and a backup interval $T_{bak}$ as 5 minutes (to balance the cost-availability tradeoff). For the {\small\texttt{large object only}} workload, we test two {\proj} configurations: the default case with backup enabled, and a case with backup disabled ({\small\texttt{without backup}}).

\vspace{-12pt}
\paragraph{Cost Savings.}
Figure~\ref{fig:docker_cost}(a) shows the accumulated monetary cost of {\proj} in comparison with an ElastiCache setup of one {\small\texttt{cache.r5.24xlarge}} Redis instance with 635.61~GB memory. By the end of hour~50, ElastiCache costs $\$518.4$, while {\proj} with {\small\texttt{all objects}} costs $\$20.52$. Caching only large objects bigger than 10~MB leads to a cost of $\$16.51$ for {\proj}. {\proj}'s pay-per-use serverless substrate effectively brings down the total cost by $96.8\%$ with a cost effectiveness improvement of $31\times$. By disabling the backup option, {\proj} further lowers down the cost to $\$5.41$, which is $96\times$ cheaper than ElastiCache. However, the low monetary cost for tenants comes at a price of impacted availability and hit ratio -- {\proj} {\small\texttt{without backup}} sees a lower hit ratio of $56.1\%$ (Table~\ref{tbl:wl}) -- thus presenting a reasonable tradeoff for tenants to choose.

{\proj}'s monetary cost is composed of three parts: (1)~serving {\small\texttt{GETs}}/{\small\texttt{PUTs}}, (2)~warming-up Lambda functions, and (3)~backing up data.  Figure~\ref{fig:docker_cost}(b)-(d) details the cost breakdown, further explaining the cost variations of different combinations of workload and {\proj} settings. 
In Figure~\ref{fig:docker_cost}(b), we see that about $41\%$ of the total cost is spent on serving data requests under the workload with {\small\texttt{all objects}}; this is because a significant portion of requests are for small objects. 
In contrast, for the {\small\texttt{large object only}} workload shown in Figure~\ref{fig:docker_cost}(c), the backup and warmup cost dominates, occupying around $88.3\%$ of the overall cost. This is because the hourly request rate for {\small\texttt{large object only}} is significantly lower than that for the {\small\texttt{all object}} workload. 
Furthermore, disabling backup leads to a dramatic cost-effectiveness improvement (see Figure~\ref{fig:docker_cost}(d)). The warm-up cost is different between Figure~\ref{fig:docker_cost}(c) and Figure~\ref{fig:docker_cost}(d), because with the backup option enabled, a warm-up invocation may trigger a backup, and thus increase the warm-up duration.

\vspace{-12pt}
\paragraph{Fault Tolerance.}
Figure~\ref{fig:docker_event} shows {\proj}'s fault tolerance activities for different cases. An object loss (losing all the replicas of more than $p$ chunks) results in a cache miss which triggers a {\small\texttt{RESET}}; the {\small\texttt{RESET}} fetches the lost object from a backing store and reinserts it into {\proj}. We observe that EC-based recovery activities and {\small\texttt{RESET}}s mostly coincide with the occurrence of request spikes at hour~15--20 and hour~34--42. Under the workload of {\small\texttt{all objects}} (Figure~\ref{fig:docker_event}(a)), we see a total of $5,720$ {\small\texttt{RESET}} events. This number is reduced to $1,085$ for the {\small\texttt{large object only}} workload (Figure~\ref{fig:docker_event}(b)), leading to an availability of $95.4\%$; as shown in Figure~\ref{fig:docker_event}(c). {\proj} {\small\texttt{without backup}} sees $3,912$ {\small\texttt{RESET}}s, which is $18.6\%$ of $21,022$ read hits in total. {\small\texttt{RESET}}s also result in a lower cache hit ratio for {\proj}, compared to ElastiCache, as shown in Table~\ref{tbl:wl}.

\begin{figure}[t]
\vspace{-5pt}
\begin{center}
\subfigure[All objects.] {
\includegraphics[width=0.21\textwidth]{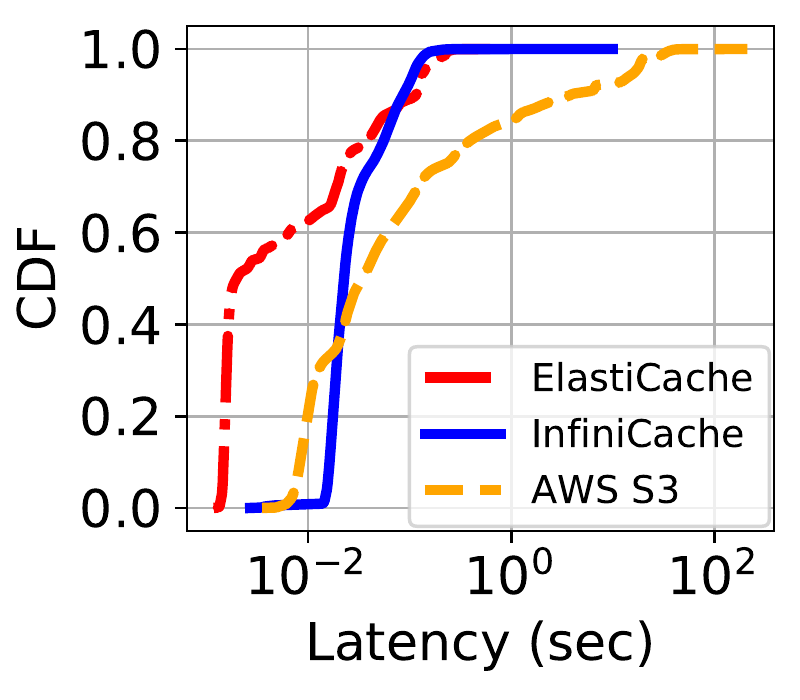}
\label{fig:all_obj}
}
\hspace{-6pt}
\subfigure[Objects $>$ 10 MB.] {
\includegraphics[width=0.21\textwidth]{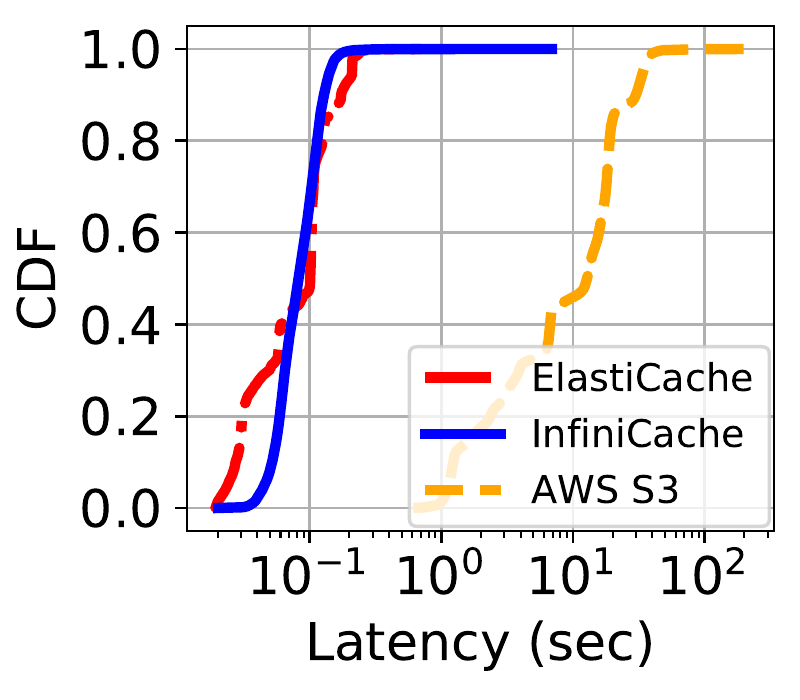}
\label{fig:large_only}
}
\vspace{-10pt}
\caption{{\proj} latencies vs. AWS S3 and ElastiCache.}
\label{fig:dockerlatency}
\vspace{-25pt}
\end{center}
\end{figure}

\begin{figure}[t]
\begin{center}
\includegraphics[width=0.36\textwidth]{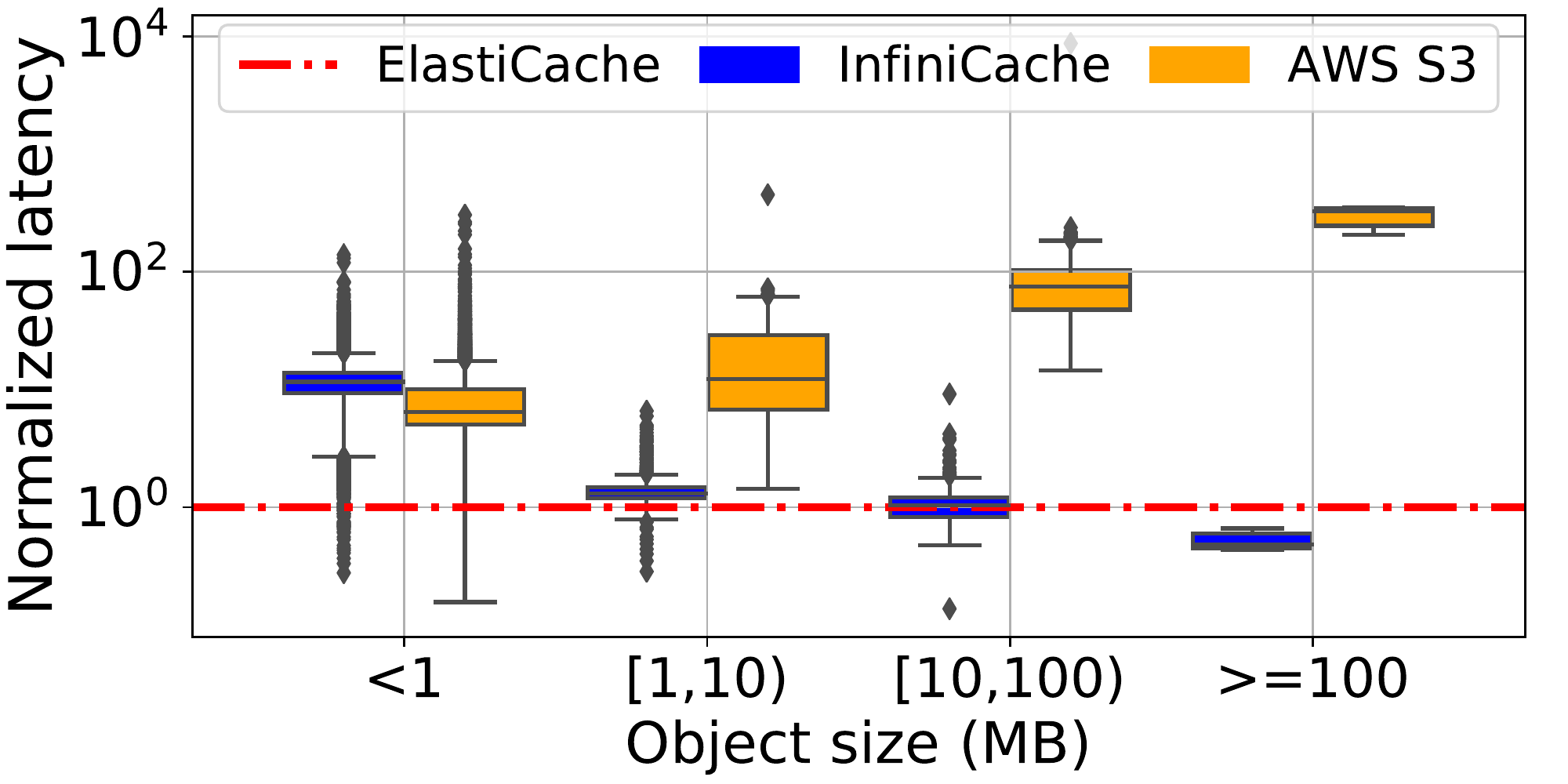}
\vspace{-10pt}
\caption{Normalized latencies grouped by object sizes.
Each is normalized to that of ElastiCache.}
\label{fig:norm_dockerlatency}
\vspace{-25pt}
\end{center}
\end{figure}

\vspace{-12pt}
\paragraph{Performance Benefit.}
We replay the first 50 hours of the Dallas trace 
against AWS S3 to simulate a deployed Docker registry service using S3 as a backing store. We compare {\proj}'s performance against AWS ElastiCache and S3  seen under the same workload ({\small\texttt{all objects}}). 
Figure~\ref{fig:dockerlatency} shows the overall trend of latency distribution, and Figure~\ref{fig:norm_dockerlatency} shows the distribution of the normalized latencies as a function of the object sizes. 

We make the following three observations.
(1)~In Figure~\ref{fig:large_only}, we see that,  compared to S3, {\proj} achieves superior performance improvement for large objects. For about $60\%$ of all large requests, {\proj} is able to achieve an improvement of at least $100\times$. This trend demonstrates the efficacy of {\proj} in serving as an IMOC in front of a cloud object store. 
(2)~{\proj} is particularly good at optimizing latencies for large objects. This is evidenced by two facts: i)~{\proj} achieves almost identical performance as ElastiCache for objects sizing from 1--100~MB; and ii)~{\proj} achieves consistently lower latencies than ElastiCache for objects larger than 100~MB (see Figure~\ref{fig:norm_dockerlatency}), due to {\proj}'s I/O parallellism. 
(3)~{\proj} incurs significant overhead for objects smaller than 1~MB (Figure~\ref{fig:norm_dockerlatency}), since fetching an object from {\proj} typically requires to invoke Lambda functions, which takes on average 13~ms and is much slower than directly fetching a small object from ElastiCache.

\vspace{-12pt}
\section{Discussion}
\label{sec:discussion}
\vspace{-6pt}

In this section, we discuss the limitations and possible future directions of {\proj}.

\begin{figure}[h]
\begin{center}
\includegraphics[width=0.38\textwidth]{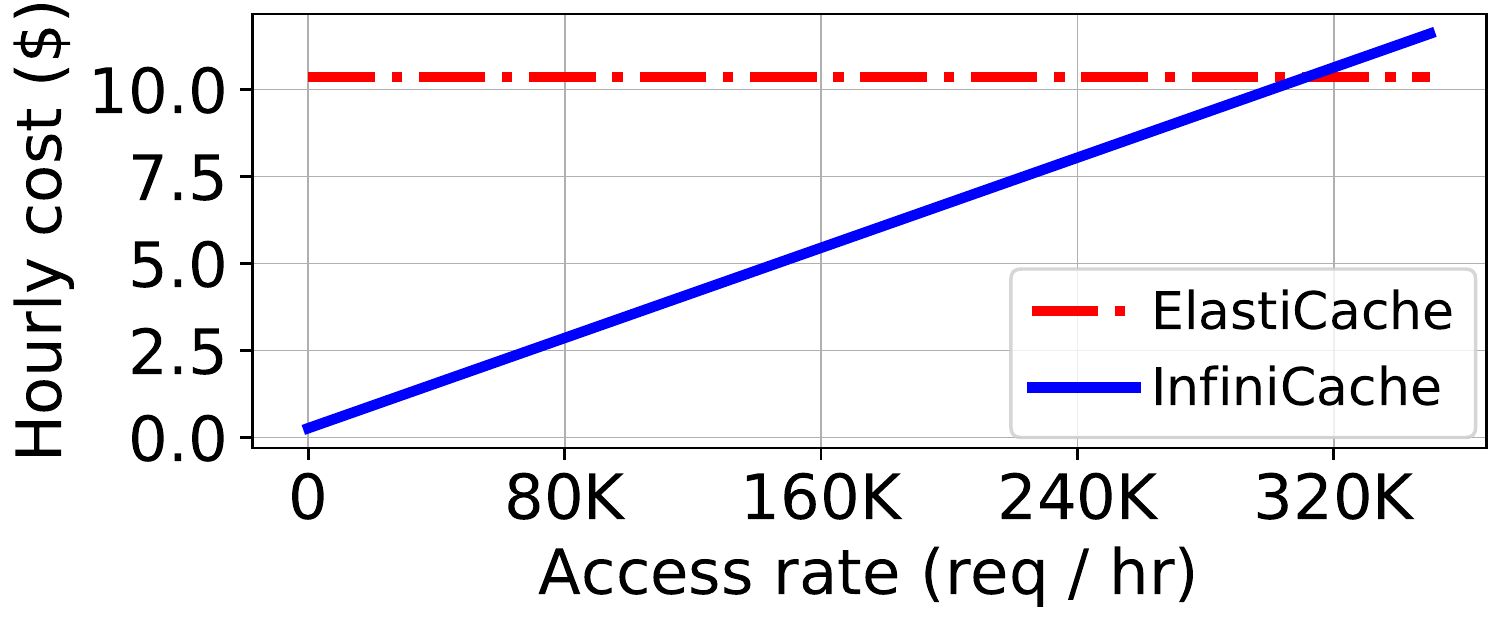}
\vspace{-10pt}
\caption{Hourly \$ cost (Y-axis) of {\proj} with 400 1.5~GB Lambdas vs. one {\small\texttt{cache.r5.24xlarge}} ElastiCache instance, as a function of access rate (X-axis).}
\label{fig:cost_rph}
\vspace{-25pt}
\end{center}
\end{figure}

\vspace{-12pt}
\paragraph{Small Object Caching.}
Small object-intensive memory caching workloads have a high traffic rate, typically ranging from thousands to hundreds of thousands of requests per second~\cite{fbkvs_sigmetrics12}. Serverless computing platforms are not cost-effective under such workloads, because the high data traffic will significantly increase the per-invocation cost and completely outweigh the pay-per-use benefit. Figure~\ref{fig:cost_rph} compares the hourly cost of {\proj} with ElastiCache, assuming the cost models in \cref{sec:data-availability-analysis} and configurations in \cref{subsec:prod}. The hourly cost increases monotonically with the access rate, and eventually overshoots ElastiCache when the access rate exceeds $312$~K requests per hour ($86$ requests per second). 

\vspace{-12pt}
\paragraph{Porting {\proj} to Other FaaS Providers.}
To the best of our knowledge, major serverless computing service providers such as Google Cloud Functions and Microsoft Azure Functions all provide function caching with various lifespans to mitigate the cost of cold startups~\cite{peeking_atc18}. Google Cloud Functions imposes similar constraints on tenants: e.g., banned in-bound TCP connections and limited function CPU/memory resources. The design of {\proj} should be portable to other major serverless computing platforms such as Google Cloud Functions, with minor source code modifications to work with Google Cloud’s APIs.

\vspace{-12pt}
\paragraph{Service Provider's Policy Changes.}
Service providers may change their internal implementations and policies in response to systems like {\proj}. On the one hand, \emph{statefulness} is urgently demanded by today's FaaS tenants -- providing durable state caching is critical to support a broader range of complex stateful applications~\cite{serverless_supercomputer_article, serverless_supercomputer_talk, crucial_middleware19} such as data analytics~\cite{locus_nsdi19} and parallel \& scientific computing~\cite{numpywren, wukong_pdsw19}. On the other hand, to strike a balance, providers could introduce new pricing models for \emph{stateful} FaaS applications -- tenants can get stateful Lambda functions by paying slightly more than that is charged by a completely stateless one. The new feature recently launched by AWS Lambda, provisioned concurrency~\cite{pconcurrency}, pins warm 
Lambda functions in memory but without any availability guarantee (provisioned Lambdas may get reclaimed, and re-initialized periodically. But the reclamation frequency is low compared to non-provisioned Lambdas), and charges tenants hourly
($\$0.015$ per GB per hour, no matter whether the provisioned functions get invoked), which is similar to EC2 VMs' pricing model.
Nonetheless, it opens up research opportunities for new serverless-oriented cloud economics. We leave developing durable storage atop {\proj} in support of new stateful serverless applications as considerations for our future work.

\vspace{-12pt}
\paragraph{Using {\proj} as a White-Box Approach.}
{\proj} presents a practical yet effective solution that exploits AWS Lambda as a black-box to achieve cost effectiveness, availability, and performance for cloud tenants. Our findings also imply that modern datacenter management systems could potentially leverage such techniques to provide short-term (e.g., intermediate data) caching for data-intensive applications such as big data analytics. Serving as a white-box solution, datacenter operators can use global knowledge to optimize data availability and locality. We hope future work will build on ours to develop new storage frameworks that can more efficiently utilize ephemeral datacenter resources.
\vspace{-12pt}
\section{Related Work}
\label{sec:related}
\vspace{-4pt}

\vspace{-5pt}
\paragraph{Cost-Effective Cloud Storage.}
Considerable prior work~\cite{racs_socc10, spanstore_sosp13, scalia_sc12, fcfs_eurosys12, costlo_nsdi15} has examined ways to minimize the usage cost of cloud storage. 
SPANStore~\cite{spanstore_sosp13} adopts a hybrid cloud approach by spreading data across multiple cloud service providers and exploits pricing discrepancies across providers. By contrast, {\proj} focuses on exploiting stateless cloud function services to achieve pay-per-use storage elasticity with dramatically reduced cost.

\vspace{-12pt}
\paragraph{Exploiting Spot Cloud Resources.}
Researchers have explored spot and burstable cloud resources to improve the cost effectiveness of applications such as memory caching~\cite{spot_eurosys17}, IaaS services~\cite{spotcheck_eurosys15}, and batch computing~\cite{spoton_socc15}. 
{\proj} differs from them in several aspects: 
(1)~ephemeral cloud functions exhibit significantly higher churn than the more stable spot instances; (2)~cloud functions are inherently ``serverless'' and cannot directly host serverful long-running applications which accept inbound network connections; and (3)~spot instances are not automatically cached by providers unlike cloud functions.

\vspace{-12pt}
\paragraph{In-Memory Key-Value Stores.}
A large body of research~\cite{mbal_eurosys15, memc3_nsdi13, distcache_fast19, masstree_eurosys12, mica_nsdi15, minos_nsdi19, kvdirect_sosp17, cliffhanger_nsdi16, zexpander_eurosys16, ramcloud_fast14} focuses on improving the performance of in-memory key-value stores for small-object intensive workloads. {\proj} is specifically designed and optimized for large objects with sizes ranging from MBs to GBs. EC-Cache~\cite{eccache_osdi16} and SP-Cache~\cite{spcache_sc18} are in-memory caches built atop Alluxio~\cite{alluxio_socc14} to provide large object caching for data-intensive cluster computing workloads. They split the large objects into smaller chunks (EC-Cache leverages erasure coding while SP-Cache directly partitions objects) and
perform curated chunk placement to achieve load balancing.
The role of erasure coding in {\proj} is multi-fold: similar to EC-Cache~\cite{eccache_osdi16}, {\proj} leverages erasure coding to mitigate the cloud functions' straggler issue; erasure coding also provides space-efficient fault tolerance against potential loss of cloud functions.

\vspace{-12pt}
\paragraph{New Applications of Serverless Computing.}
Researchers have identified new applications
for serverless computing in data analytics~\cite{pywren_socc17, wukong_pdsw19}, video processing~\cite{excamera_nsdi17, sprocket_socc18}, linear algebra~\cite{numpywren}, machine learning~\cite{cirrus_socc19, dl_serving_ic2e18}, and software compilation~\cite{gg_atc19}.
However, these applications exploit the computing power of serverless platforms to parallelize and accelerate compute-intensive jobs, whereas {\proj} presents a completely new use case of cloud function services---implementing a stateful storage service atop stateless cloud functions by exploiting transparent function caching.

\vspace{-12pt}
\section{Conclusion}
\label{sec:conclusion}
\vspace{-6pt}

With web applications becoming increasingly storage-intensive, it is imperative to revisit the design of in-memory object caching in order to efficiently deal with both small and large objects. We have presented a novel in-memory object caching solution that achieves high cost effectiveness and good availability for large object caching by building {\proj} on top of a popular serverless computing platform (AWS Lambda). For the first time in the literature, {\proj} enables request-driven pay-per-use elasticity at the cloud storage level with a serverless architecture. {\proj} does this by synthesizing a series of techniques including erasure coding and a delta-sync-based data backup scheme. Being serverless-aware, {\proj} intelligently orchestrates ephemeral cloud functions and improves cost effectiveness by $31\times$ compared to ElastiCache, while maintaining $95.4\%$ availability for each hour time window.
{\proj}'s source code is available at: 
\begin{center}
\vspace{-5pt}
\url{https://github.com/mason-leap-lab/InfiniCache}.
\end{center}





\label{startofrefs}
\clearpage
\newpage
\balance

\vspace{-12pt}
\section*{Acknowledgments}
\vspace{-8pt}

We are grateful to our shepherd, Carl Waldspurger, as well as the anonymous reviewers, for their valuable comments and suggestions that significantly improved the paper. We would also like to thank Benjamin Carver and Richard Carver for their careful proofreading. This work is sponsored in part by NSF under CCF-1919075, CCF-1756013, IIS-1838024, and AWS Cloud Research Grants.

{
\bibliographystyle{plain}
\bibliography{refs}
}

\end{document}
